\documentclass[namedreferences]{solarphysics}
\usepackage[hyperref,optionalrh,solaromanenum]{spr-sola-addons} 
\usepackage{graphicx}                    
\usepackage{color}                       
\usepackage{breakurl}                         
\usepackage{savesym}
\savesymbol{iint}
\savesymbol{iiint}
\usepackage{amsmath}
\usepackage{sidecap}
\usepackage{multirow}

\begin{document}

\begin{article}

\begin{opening}

\title{Forbush Decrease Prediction Based on the Remote Solar Observations}

%
\author{M.~\surname{Dumbovi\'c}$^{1}$\sep
        B.~\surname{Vr\v snak}$^{1}$\sep
        J.~\surname{\v Calogovi\'c}$^{1}$
       }
       
%
\runningauthor{\textit{Dumbovi\'c et al.}}
\runningtitle{\textit{Forbush Decrease Prediction Based on the Remote Solar Observations}}

%
 \institute{$^{1}$ Hvar Observatory, Faculty of Geodesy, University of Zagreb, Zagreb, Croatia;
                     email: \url{mdumbovic@geof.hr}\\ 
           }


\begin{abstract}
We employ remote observations of coronal mass ejections  (CMEs) and the associated solar flares to forecast the CME-related Forbush decreases, \textit{i.e.}, short-term depressions in the galactic cosmic-ray flux. The relationship between the Forbush effect at the Earth and remote observations of CMEs and associated solar flares is studied \textit{via} a statistical analysis. Relationships between Forbush decrease magnitude and several CME/flare parameters was found, namely the initial CME speed, apparent width, source position, associated solar-flare class and the effect of successive-CME occurrence. Based on the statistical analysis, remote solar observations are employed for a Forbush-decrease forecast. For that purpose, an empirical probabilistic model is constructed that uses selected remote solar observations of CME and associated solar flare as an input, and gives expected Forbush-decrease magnitude range as an output. The forecast method is evaluated using several verification measures, indicating that as the forecast tends to be more specific it is less reliable, which is its main drawback. However, the advantages of the method are that it provides early prediction, and that the input is not necessarily spacecraft-dependent.
\end{abstract}

%

\end{opening}

\section{Introduction}
\label{intro} 

Short-term depressions in the galactic cosmic ray (GCR) flux were first observed by \inlinecite{forbush37} and \inlinecite{hess37} and termed ``Forbush decreases" (FD). There are two types: one caused by corotating interaction regions (see, \textit{e.g.}, \opencite{richardson04}) and the other caused by interplanetary counterparts of CMEs (ICMEs) and/or the shocks that they drive (see, \textit{e.g.}, \opencite{cane00} and \opencite{belov09}). ICME-related FDs are typically asymmetric, and they have duration of a few days, magnitude larger than the daily CR-flux variations, and an onset close to the arrival of an ICME. If an ICME is preceded by a shock/sheath region, a characteristic two-step FD is expected, the first step coming from the sheath, whereas the second depression is associated with the ejecta (for an overview on ICME-related FDs see, \textit{e.g.}, \opencite{cane00} and \opencite{richardson10}). However, two-step FDs are not very common and FDs generally show a diverse and complex structure, even in the case of single ICMEs (\opencite{jordan11}).

The physical mechanism behind the modulation of cosmic rays can be described in general by a transport equation (\opencite{parker65}), which combines contributions from convection and adiabatic energy loss by a fast stream and enhanced drift and scattering properties of strong and fluctuating magnetic field. These effects were considered in several models trying to explain FDs (\textit{e.g.} \opencite{leroux91,cane95,wibberenz97,wibberenz98,krittinatham09,kubo10}). Since the models are based on the interaction of charged particles with interplanetary shock/sheath region and/or ejecta, any implementation of such models for forecasting purposes would be dependent on the observed interplanetary ICME characteristics. There have been many observational studies as well, relating FD magnitude to ICME interplanetary properties such as magnetic-field strength and fluctuations, and ICME speed (see, \textit{e.g.}, \opencite{chilingarian10,richardson11a,dumbovic12b,blanco13a} and references therein). The good relationship between \textit{in-situ} properties of ICMEs and FDs enables using real-time near-Earth \textit{in-situ} measurements to forecast the ICME-related Forbush effect. However, given the current position of spacecraft providing such measurements, predictions can precede the FD event only about one hour in advance. Furthermore, a recent study by \inlinecite{thomas15} provided strong evidence for the modulation of GCR flux by remote solar-wind structures, \textit{i.e.} this indicates that Forbush decreases can even be produced remotely.

To obtain timely information on a Forbush decrease, a forecast method is needed that can derive interplanetary properties of ICMEs based on the remote solar observations of CMEs and associated solar phenomena (\textit{e.g.} solar flares or EUV dimmings). Moreover, since FD magnitude is dependent on the magnetic field and speed of the ICME, these should be derived from the initial CME properties during the liftoff. However, neither the magnetic field nor CME initial speed are directly observable. \inlinecite{chertok13} used the magnetic flux at the photospheric level beneath EUV dimmings and post-eruption arcades associated with CMEs as a measure of a CME magnetic field and obtained a good correlation with the FD magnitude. However, only a fraction of CMEs are associated with EUV dimmings and moreover, there are CMEs without any chromospheric or low coronal signatures (stealth CMEs, see, \textit{e.g.}, \opencite{robbrecht09,howard13} and references therein). Recent studies have shown that using white-light coronagraphic observations of CMEs can relate their properties to FD magnitudes. FD magnitude was found to be larger for faster CMEs (\opencite{blanco13a,belov14}), CMEs with larger apparent width (\opencite{kumar14,belov14}) and CMEs with greater mass (\opencite{belov14}). In addition, \inlinecite{belov09} found that the sources of the largest Forbush effects are usually located in the central part of the visible solar disc. However, it should be noted that CME measurements suffer from projection effects and can therefore be taken only as proxies of CME initial coditions. In addition, the relations between these observed CME properties and FDs are much weaker compared to ICME--FD relations.

\section{Data and Method for Statistical Analysis}
\label{data}
 
We aim to relate remote CME observational properties to FD magnitude and employ statistical relationships in forecasting. In the article by \inlinecite{dumbovic15a} an analogous problem was considered, trying to statistically relate remote CME observational properties to geomagnetic-storm strength. Therefore, we use the same methodology here. By analogy with the geoeffectiveness of CMEs, as a measure of geomagnetic response we adopt the term GCR-effectiveness (used by \opencite{kumar14}) as a measure of the cosmic-ray response.

We use a sample of events listed by \inlinecite{dumbovic15a} and supplement it with FD events. The original list contains 211 CMEs with associated solar flares and geomagnetic response measured by the Disturbance storm time (Dst) index. The list consists of various CME observational properties taken from the \textit{Solar and Heliospheric Observatory} (SOHO) \textit{Large Angle and Spectrometric Coronagraph} (LASCO) CME Catalog (\opencite{yashiro04}: \url{cdaw.gsfc.nasa.gov/CME list/}) and solar flare data taken from the National Oceanic and Atmospheric Administration (NOAA) X-ray solar flare list (\url{ftp://ftp.ngdc.noaa.gov/STP/space-weather/solar-data/solar-features/solar-flares}). Each CME/flare event in the list is  associated with a Dst value and time (time of the reported/measured Dst anomaly). For each CME in the list there is also an ``interaction parameter" describing the likelihood that the CME interacts with some other CME(s). It is derived using CME timing, width, and source position with respect to other ``close" CMEs (for more details see \opencite{dumbovic15a}). The interaction parameter [$i$] is a discrete parameter that can have four different values, corresponding to four levels of ``interaction probability":\\

\noindent $i=1$: ``SINGLE" (S) events -- no interaction;\\
$i=2$: ``SINGLE?" (S?) events -- interaction not likely;\\
$i=3$: ``TRAIN?" (T?) events -- probable interaction;\\
$i=4$ ``TRAIN" (T) events -- interaction highly probable.\\

For each event on the list we searched for a corresponding response in the relative pressure-corrected cosmic-ray (CR) count in the ground-based neutron monitor (NM) data taken from the \textit{Space Physics Interactive Data Resource} (SPIDR, \url{spidr.ngdc.noaa.gov/spidr/}). We searched the time interval spanning 5 days before and 15 days after the reported Dst anomaly to find a response in the CR count (if there is one). To reduce the effect of daily variations we used the average of three to four different mid-latitude NM stations (depending on data availability) at different asymptotic longitudes, but of similar rigidity (Novosibirsk, Calgary, Kiel, and Magadan, with vertical cutoff rigidity 2.91 GV, 1.09 GV, 2.29 GV, and 2.10 GV, respectively; for method details see \opencite{dumbovic11}). This method reduces the daily variations, but does not remove them completely. Therefore, a threshold of $1$\,\% (comparable to the daily variation amplitude) is chosen to distinguish GCR-effective events. If a clear depression in the CR count with a magnitude $>$1\,\% (from the onset point to the time of maximum decrease) is observed around the reported Dst timing (\textit{i.e.} the time of the minimum Dst is within the FD duration interval), the event is regarded as GCR-effective; otherwise it is not regarded as GCR-effective. Thus we treat small FDs ($<$1\,\%) in the same way as ``missing" FDs (\textit{i.e.} when there is no event because the CME did not arrive at the Earth). The CR counts were normalized to the CR count in the quiet period before any disturbance. In some cases, where two consecutive geomagnetic storms could be identified separately, but only one Forbush decrease is observed, the two events are merged into one event, which is then regarded as an interacting-CMEs event (interaction parameter, $i=4$). In such cases the CMEs involved are treated as one event, characterized by the solar parameters of the fastest CME involved in the interaction and with the apparent width of the widest CME involved. This, in addition to data gaps for several events, resulted in a new list of 187 CME--flare--Dst--FD associations.

By analogy with the method used by \inlinecite{dumbovic15a}, FD magnitudes [$FD$] were grouped into four different levels of GCR-effectiveness:\\ 

\noindent $FD<1\,\%$ (not GCR-effective);\\
$1\%<FD<3\,\%$ (moderately GCR-effective);\\
$3\%<FD<6\,\%$ (strongly GCR-effective);\\
$FD>6\,\%$ (intensively GCR-effective).\\

We focus on specific CME/flare parameters \textit{viz.} the initial CME speeds and angular width, solar flare soft X-ray class and location, and interaction parameter. To relate these parameters to $FD$ a statistical analysis was performed using $FD$-distributions as a statistical tool. The selected CME/flare parameters were also binned. For some parameters the binning was obvious (\textit{e.g.} interaction parameter) as they are already discrete parameters. For continuous parameters, all of the bins have approximately the same number of events, therefore, these bins are not equidistant. The $FD$-distribution mean is then calculated, which can be correlated with the change in the mean value of the (discrete) CME/flare parameter. To support/substantiate our analysis we use the method of overlapping bins. With this method, in addition to the original binning, an alternative binning is used and the results for both are then compared. The benefit is twofold: firstly, if the alternative binning leads to the same results as the original binning, it contributes to the plausability of the results; secondly, in this way additional data points are obtained for the correlation of the $FD$-distribution mean and the mean value of the (discrete) CME/flare parameter. Both original and alternative bins, as well as the corresponding number of the events are given in Table \ref{tab1}.

%
\begin{table}
\caption{CME/flare parameters binning and corresponding number of events}
\label{tab1}
\begin{tabular}{ccccc}
\cline{1-5}
					&	\multicolumn{2}{c}{ } 						 & \multicolumn{2}{c}{ }\\
CME/flare & \multicolumn{2}{c}{original bins} & \multicolumn{2}{c}{alternative bins}\\
\cline{2-5}
parameter & 	  & Number 		&  		 & Number\\
					&	bin	& of events	&	bin	 & of events\\
\cline{1-5}
CME      &                                  &                & $<60^{\circ}$                    & 20\\
apparent & $<120^{\circ}$                   & 56             & $60^{\circ}$\,--\,$100^{\circ}$  & 22\\
width    & $120^{\circ}$\,--\,$360^{\circ}$ & 32             & $100^{\circ}$\,--\,$140^{\circ}$ & 25\\
$[w]$    & $360^{\circ}$                    & 99             & $140^{\circ}$\,--\,$360^{\circ}$ & 21\\
         &                                  &                & $360^{\circ}$                    & 99\\
\cline{1-5}
        & 400\,--\,650 $\mathrm{km\,s^{-1}}$   & 31 &                                      &   \\
CME     & 650\,--\,800 $\mathrm{km\,s^{-1}}$   & 27 & 400\,--\,700 $\mathrm{km\,s^{-1}}$   & 40\\
initial & 800\,--\,1000 $\mathrm{km\,s^{-1}}$  & 31 & 700\,--\,1000 $\mathrm{km\,s^{-1}}$  & 49\\
speed   & 1000\,--\,1200 $\mathrm{km\,s^{-1}}$ & 33 & 1000\,--\,1500 $\mathrm{km\,s^{-1}}$ & 50\\
$[v]$   & 1200\,--\,1700 $\mathrm{km\,s^{-1}}$ & 34 & $>1500$ $\mathrm{km\,s^{-1}}$        & 48\\
        & $>1700$ $\mathrm{km\,s^{-1}}$        & 31 &                                      &   \\
\cline{1-5}
CME/flare       & $<0.35$ R$_{\odot}$           & 28  &                                &   \\
source position & $0.35$\,--\,$0.5$ R$_{\odot}$ & 30  & $<0.45$ R$_{\odot}$            & 49\\
distance from   & $0.5$\,--\,$0.6$ R$_{\odot}$  & 25  & $0.45$\,--\,$0.65$ R$_{\odot}$ & 44\\
the center      & $0.6$\,--\,$0.75$ R$_{\odot}$ & 34  & $0.65$\,--\,$0.85$ R$_{\odot}$ & 41\\
of the solar    & $0.75$\,--\,$0.9$ R$_{\odot}$ & 32  & $>0.85$ R$_{\odot}$            & 53\\
disc $[r]$      & $>0.9$ R$_{\odot}$            & 38  &                                &   \\
\cline{1-5}
             &                                    &    & $<2.5\cdot10^{^-6}$Wm $^{^-2}$          & 28\\
solar flare  & $<10^{-5}$Wm $^{^-2}$              & 84 & $2.5$\,--\,$5\cdot10^{^-6}$Wm $^{^-2}$  & 33\\
soft X-ray   & $10^{-5}$\,--\,$10^{-4}$Wm$^{^-2}$ & 67 & $5$\,--\,$15\cdot10^{^-6}$Wm $^{^-2}$   & 34\\
flux peak    & $>10^{-4}$Wm $^{^-2}$              & 36 & $15$\,--\,$50\cdot10^{^-6}$Wm $^{^-2}$  & 34\\
value $[I]$  &                                    &    & $50$\,--\,$150\cdot10^{^-6}$Wm $^{^-2}$ & 33\\
             &                                    &    & $>150\cdot10^{^-6}$Wm $^{^-2}$          & 25\\
\cline{1-5}
interaction & $i=1$ & 83 & $i=1$ \& $i=2$ & 108\\
parameter   & $i=2$ & 25 & $i=2$ \& $i=3$ & 46\\
$[i]$       & $i=3$ & 21 & $i=3$ \& $i=4$ & 79\\
            & $i=4$ & 58 &                &   \\
\cline{1-5}
\end{tabular}
\end{table}

\section{Statistical Analysis}
\label{statistics}

%
\begin{figure}
\centerline{\includegraphics[width=0.47\textwidth]{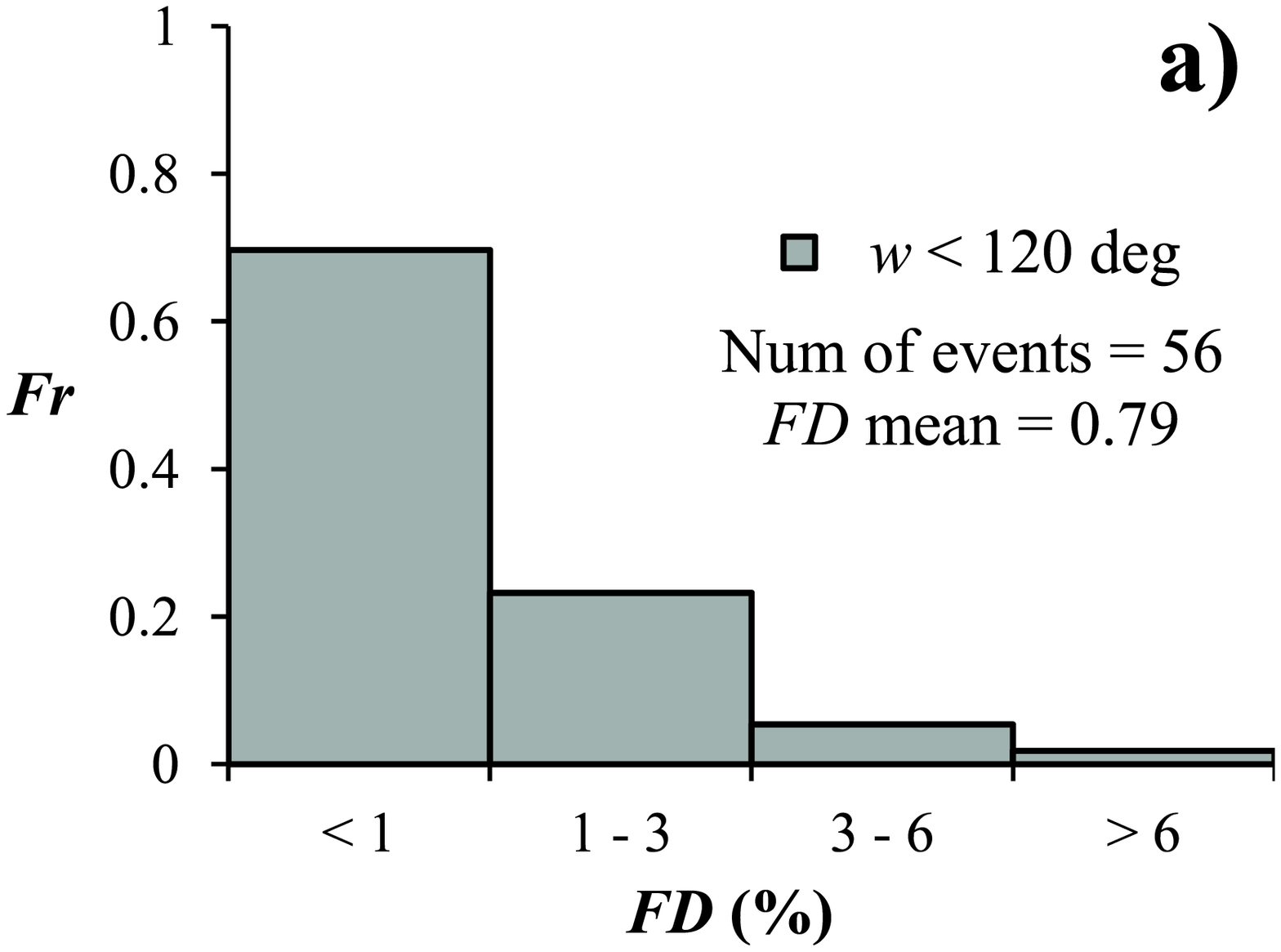}
						\hspace{0.03\textwidth}
						\includegraphics[width=0.47\textwidth]{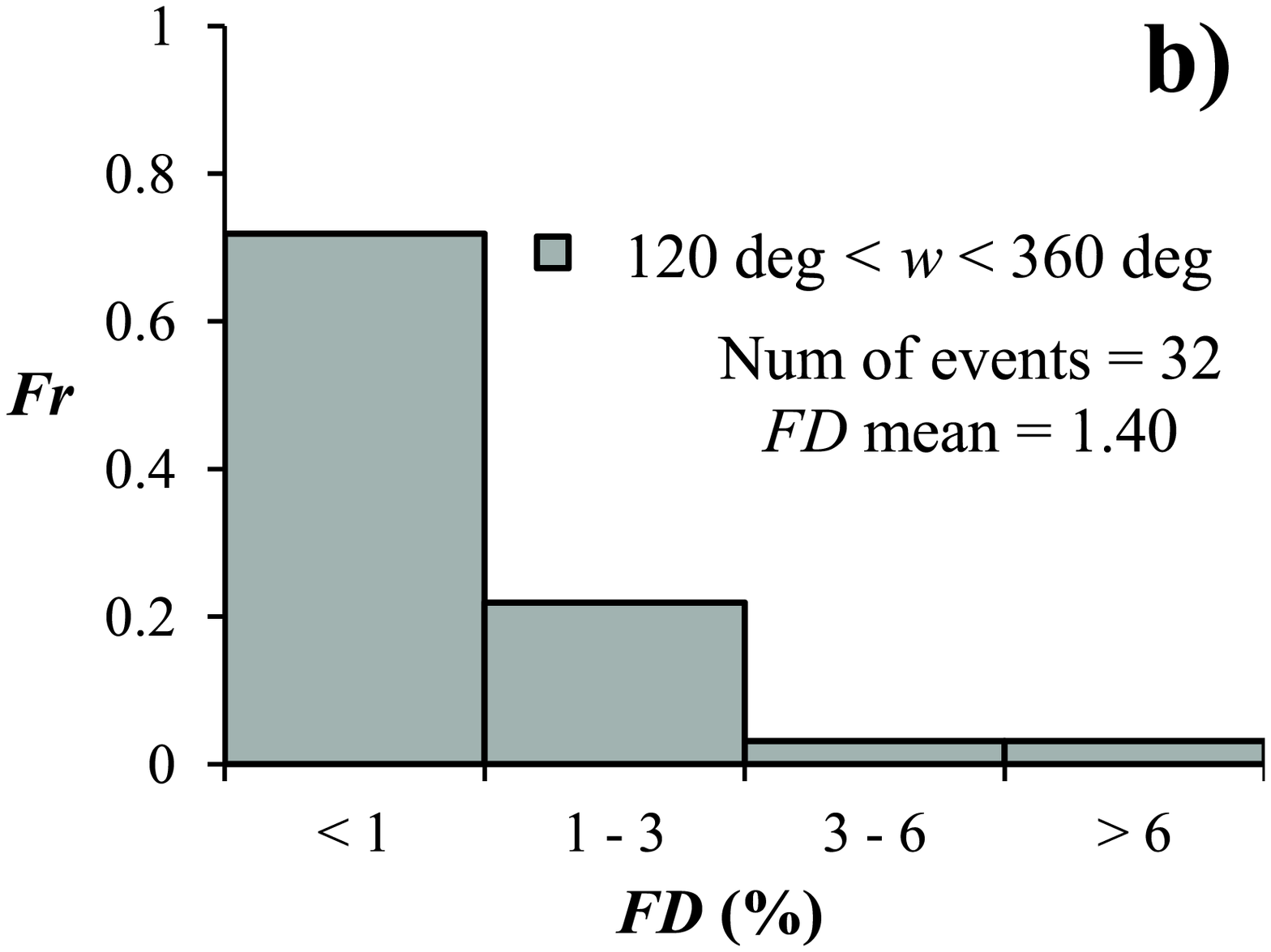}
						}
\vspace{0.03\textwidth}
\centerline{\includegraphics[width=0.47\textwidth]{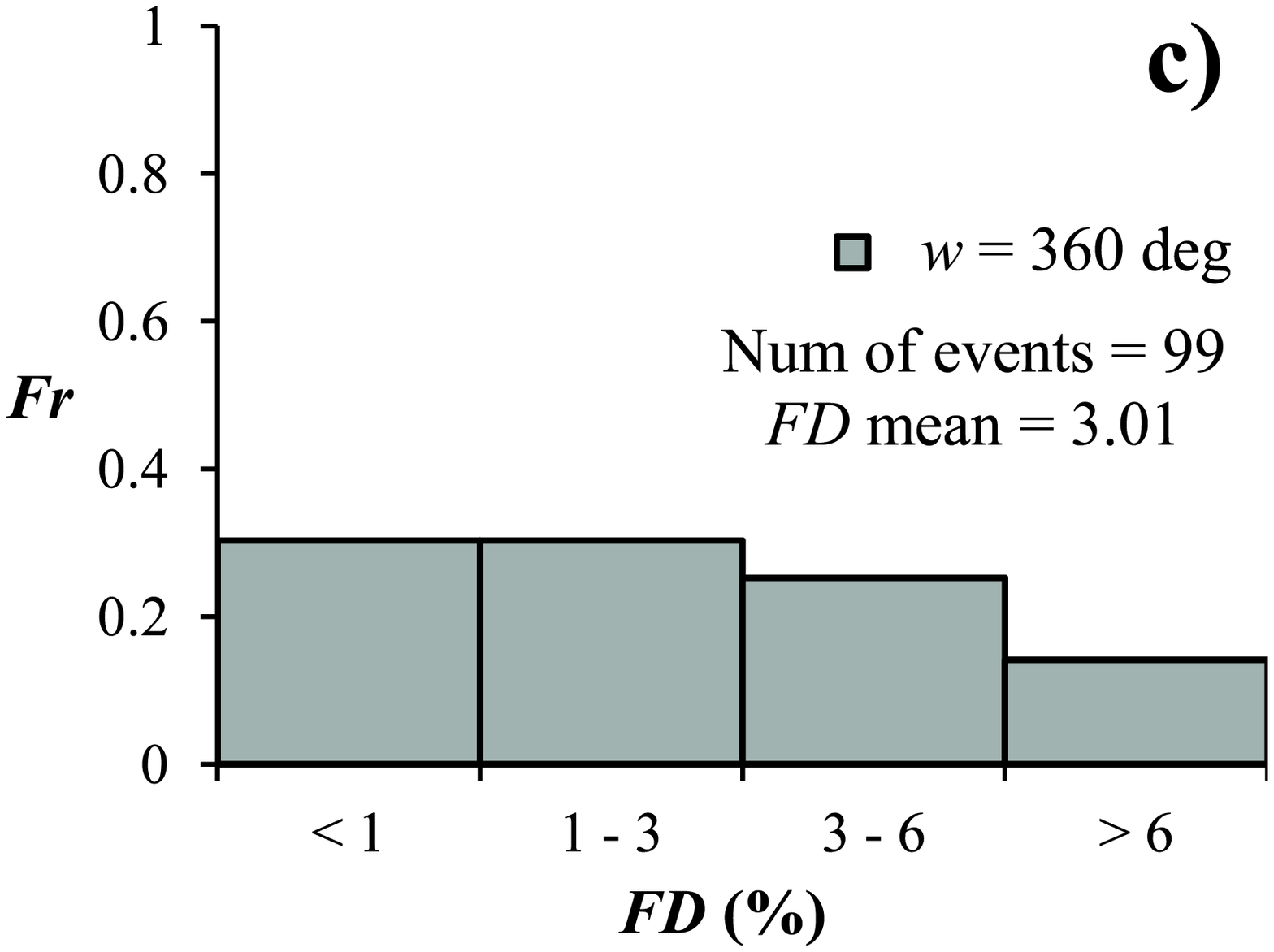}
						\hspace{0.03\textwidth}
						\includegraphics[width=0.47\textwidth]{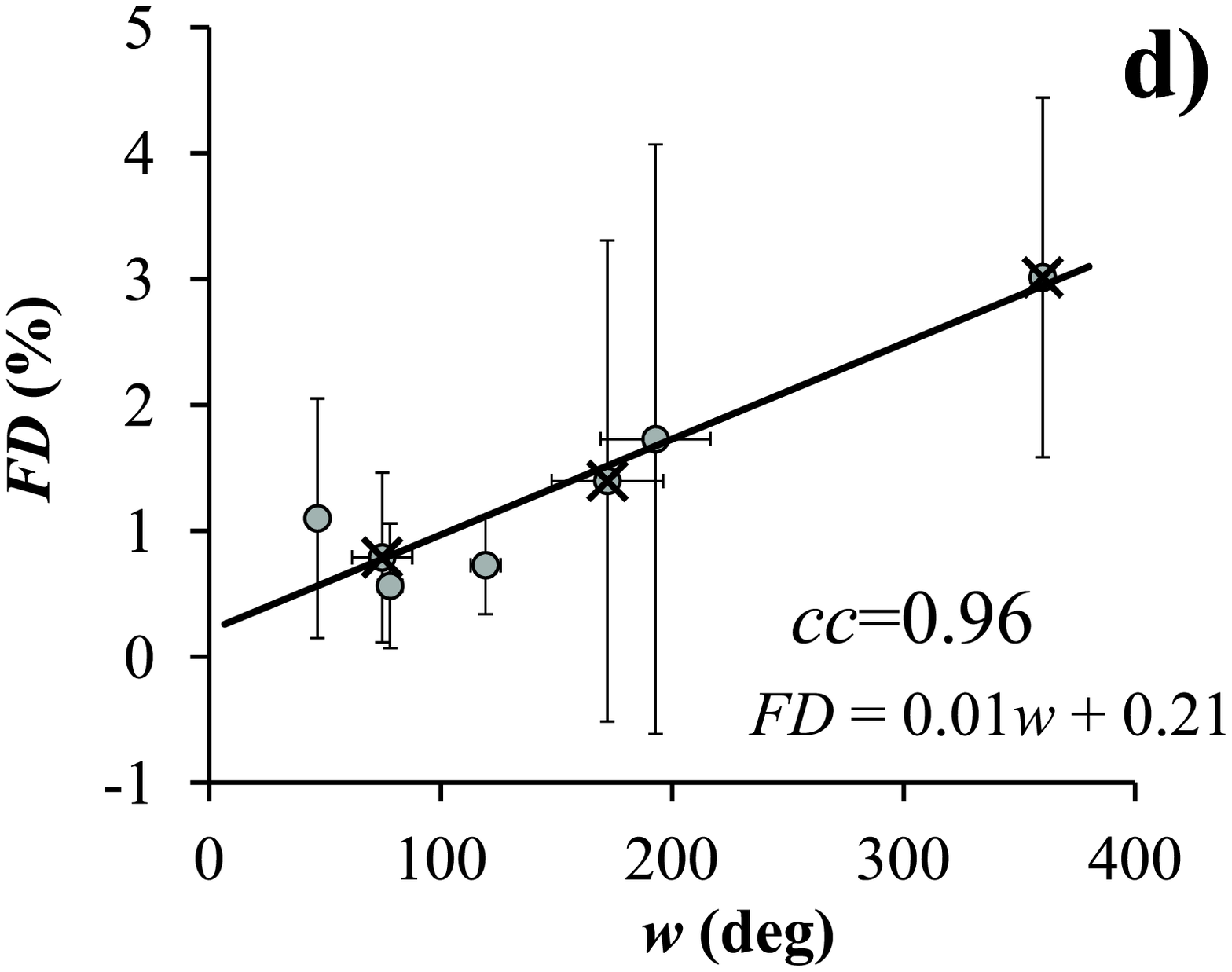}
}
\caption{$FD$ relative frequencies $[Fr]$ for different $FD$ and CME apparent-width $[w]$ bins (a--c), and $FD$ distribution mean as a function of the bin-averaged value for the CME apparent width (d). A linear fit to all of the data obtained by the method of overlapping bins is given in d with fitting parameters and a correlation coefficient $[cc]$. The data corresponding to original bins (used for distributions in a--c) are marked by crosses. Standard deviation is given by the error bars.}
\label{fig1}
\end{figure}

The first parameter analyzed is the CME apparent width. The events in our data set were first categorized into three different CME apparent width bins, following the categorization from the SOHO/LASCO CME catalog into non-halo, partial halo, and halo CMEs (original bins for $w$ are defined in Table \ref{tab1}). Due to the fact that (possibly) interacting CMEs are regarded as one entity (``TRAIN" and ``TRAIN?" events, see Section \ref{data}), these events were associated with the width of the widest CME involved in (possible) interaction. Using $FD$ binning for GCR-effectiveness explained in Section \ref{data}, three $FD$ distributions were made (Figure \ref{fig1}a--c). The mean of each distribution was calculated, as well as the average value of width within a certain range, to quantitatively examine changes in the $FD$ distribution for different CME width ranges (presented by crosses in Figure \ref{fig1}d). The whole procedure was repeated with alternative binning, using the apparent widths listed in the SOHO/LASCO CME catalog (alternative bins for $w$ are defined in Table \ref{tab1}). A linear-least-square fit to all of the data in Figure \ref{fig1}d (for both original and alternative bins) shows a strong correlation. However, large standard deviations can be seen, \textit{i.e.} a large scatter of $FD$ values is present within each bin. This indicates that CMEs with larger apparent width are in general more GCR-effective, as concluded previously by \inlinecite{kumar14} and \inlinecite{belov14}, but there is a large event-to-event variability. We expect to observe this large variability for each of the solar parameters because although FD magnitude was observed to be related to some of the solar parameters, a strong correlation was not found (unless average values are used, as in Figure \ref{fig1}d). This implies a complex relation between the FD magnitude and a number of possible parameters, and therefore motivates a probabilistic approach.  

\begin{figure}
\centerline{\includegraphics[width=0.47\textwidth]{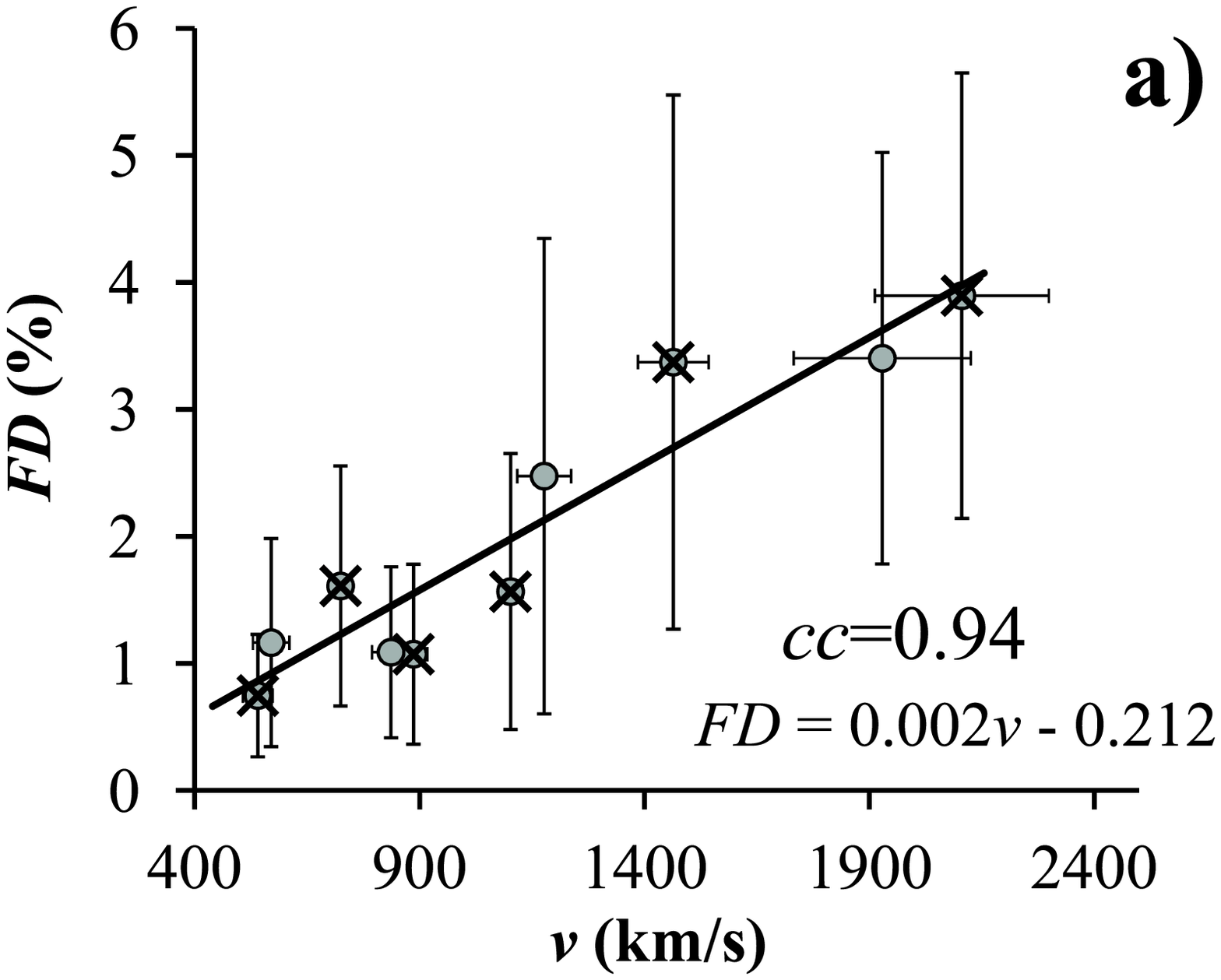}
						\hspace{0.03\textwidth}
						\includegraphics[width=0.47\textwidth]{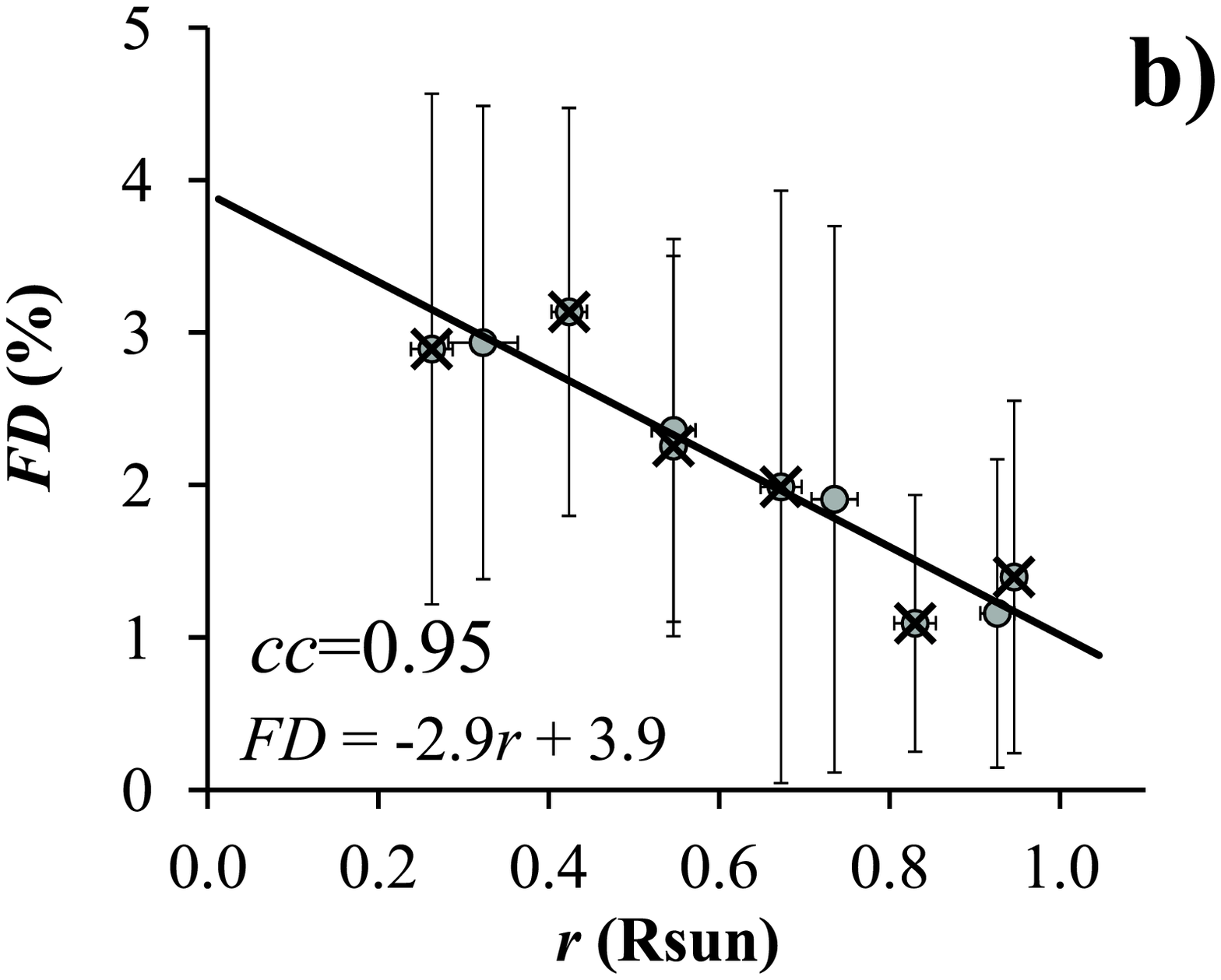}
						}
\vspace{0.03\textwidth}
\centerline{\includegraphics[width=0.47\textwidth]{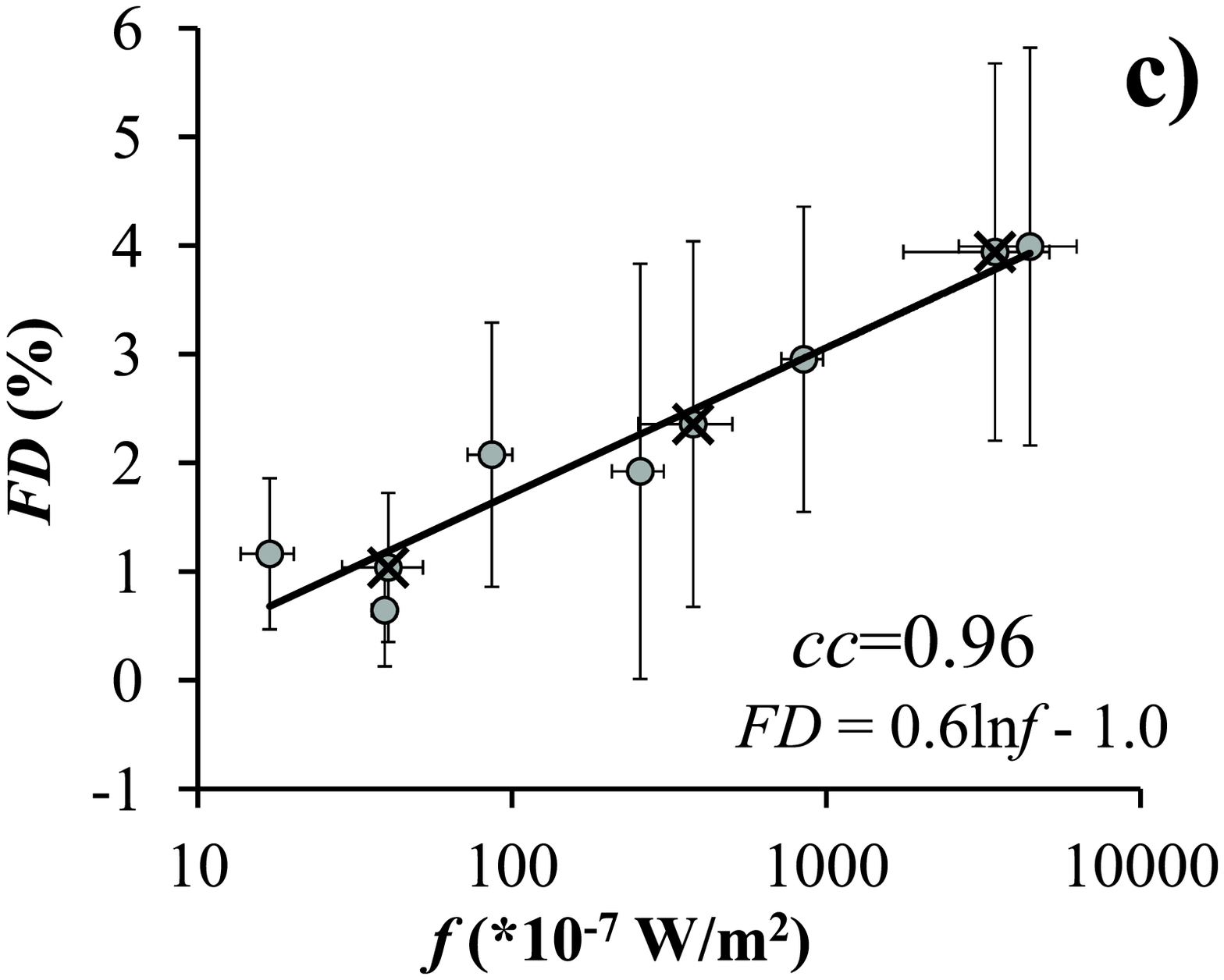}
						\hspace{0.03\textwidth}
						\includegraphics[width=0.47\textwidth]{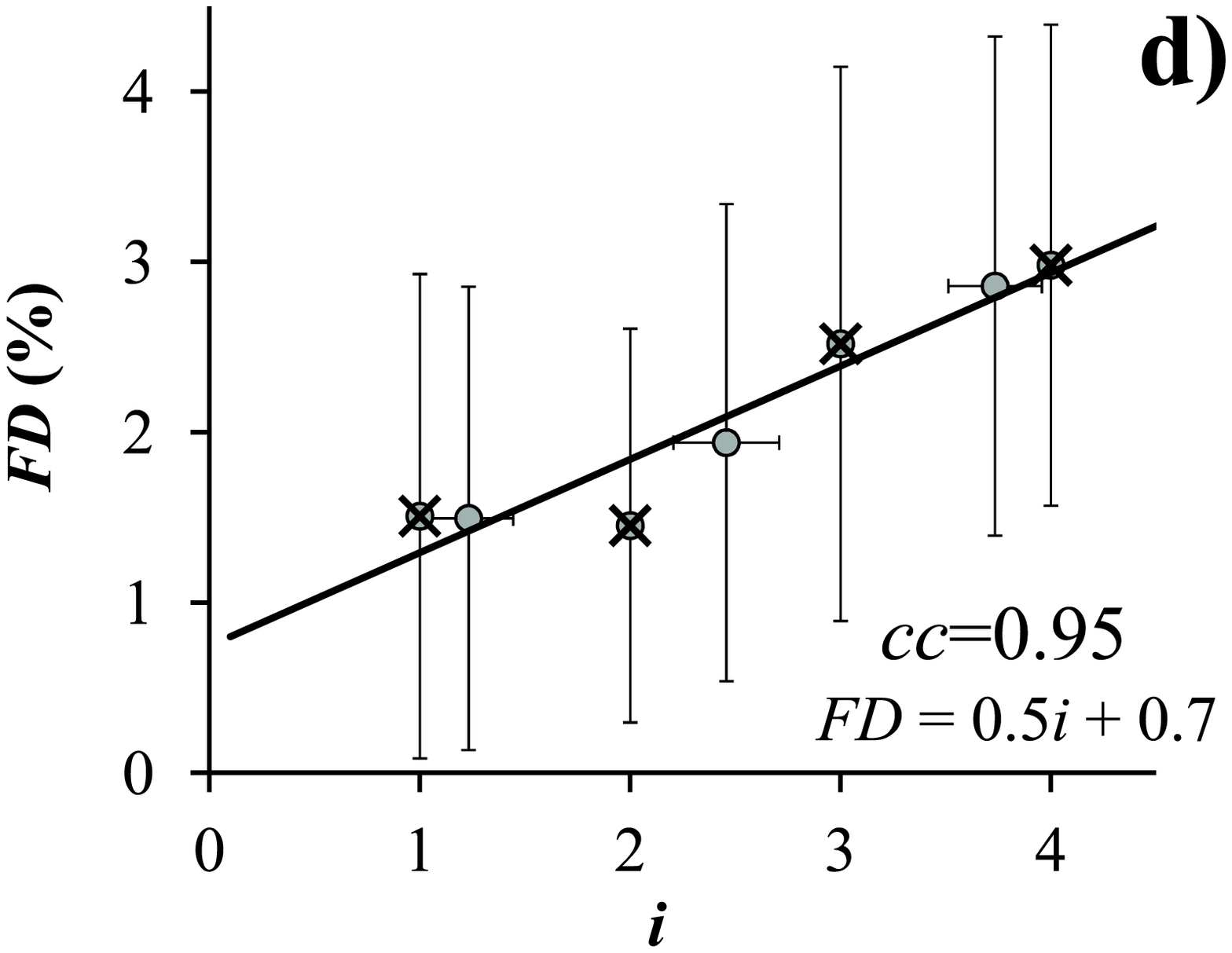}
}
\caption{$FD$ distribution mean as a function of the bin-averaged value for the CME initial speed $[v]$ (a), CME/flare source position $[r]$ (b), solar flare soft X-ray peak intensity (logarithmic scale) $[f]$ (c), and CME--CME interaction level $[i]$ (d). Best-fit to all of the data obtained by the method of overlapping bins is given for each of the solar parameter, as well as the corresponding fitting parameters and a correlation coefficient $[cc]$. The data corresponding to original bins are marked by crosses. Standard deviations are given by the error bars.}
\label{fig2}
\end{figure}

Next, we analyze the first-order (linear) CME speed $[v]$ derived from LASCO-C2 and -C3 images. The events in our data set were categorized into six different CME speed bins (original bins are listed in Table \ref{tab1}). For each category of $v$ we obtain the GCR-effectiveness distribution, similarly as for apparent width $[w]$. For each distribution the distribution mean and corresponding average value of speed within a certain range is calculated (shown by crosses in Figure \ref{fig2}a). We again apply the method of overlapping bins (alternative bins are listed in Table \ref{tab1}). A linear-least-square fit to all of the data in Figure \ref{fig2}a (for both original and alternative bins) shows a strong correlation, indicating that FD magnitude is larger for faster CMEs, in agreement with \inlinecite{blanco13a} and \inlinecite{belov14}, although large standard deviations again imply large event-to-event variability.

Similar analysis, using the method of overlapping bins, was repeated for the CME/flare source position, \textit{i.e.} for the radial distance of the CME/flare source position from the center of the solar disc, expressed in solar radii $[r]$. The first binning of $r$ (shown by crosses in Figure \ref{fig2}b) contains six bins, whereas the alternative binning results in four bins (original and alternative bins for $v$ are shown in Table \ref{tab1}). The linear-least-square fit to all of the data in Figure \ref{fig2}b results in a strong correlation, but with large standard deviations. Nevertheless, $FD$ is found to be related to the CME/flare source position, namely it is found that $FD$ is larger for CMEs/flares originating closer to the center of the solar disc, in agreement with \inlinecite{belov09}.

%

FD magnitude was related to associated flare strength using the soft X-ray flux peak value $[f]$ grouped according to the widely used classification of soft X-ray flares into B-, C-, M-, and X-class flares (original bins for $[f]$ are shown in Table \ref{tab1}). Due to the small number of B-class flares, they share a bin with C-class flares. The alternative binning with six different ranges of soft X-ray flux peak value was applied, where each bin contains approximately the same number of events (alternative bins for $I$ are shown in Table \ref{tab1}). The best fit to all of the data obtained by the method of overlapping bins, in spite of the large standard deviations within bins, reveals a logarithmic dependence (Figure \ref{fig2}c), where $FD$ is found to be larger for stronger flares.

Finally, we relate the FD magnitude $[FD]$ to the CME--CME interaction parameter, which describes the likelihood that the CME will interact with another CME. As described in Section \ref{data}, this is a discrete, dimensionless parameter (original bins for $i$ are shown in Table \ref{tab1}). For the purpose of the overlapping-bins method, alternative binning was applied, having three different bins, where events with close interaction parameter values were put in the same bin (\textit{i.e.} bin 1 containing $i=1$ and $i=2$ events, bin 2 containing $i=2$ and $i=3$ events, bin 3 containing $i=3$ and $i=4$ events; see the alternative bins for $i$ in Table \ref{tab1}). A linear-least-square fit to all of the data in Figure \ref{fig2}d (for both original and alternative bins) shows a strong correlation. As with all of the previous solar parameters, the standard deviations are again large within the bins. However, the results indicate that $FD$ is generally related to the interaction parameter, \textit{i.e.} that $FD$ is larger for interacting/multiple CMEs.

\section{Empirical Statistical Model for Predicting Forbush Decrease Magnitude}
\label{model}

We use the results of the statistical analysis presented in the previous section to construct the distribution of FD magnitude for a specific set of remote solar observations of a CME and associated flare. The procedure is analogous to the one described by  \inlinecite{dumbovic15a}.

As a mathematical tool we use the shifted geometric distribution (\textit{i.e.} the geometric distribution for the random variable $Y=X-1$, see, \textit{e.g.}, \opencite{stirzaker}):


\begin{equation}
P(X=k)= p(1-p)^{k}\,,
\label{eq1}
\end{equation}

\noindent
where $P(X=k)$ is the probability that there will be $k$ trials with a failure before the first trial with a success, and $p$ is the probability of the success in each trial ($k=0,1,2,...$ is the number of trials). The shifted geometric distribution can easily be constructed if the distribution mean $[m]$ is known:


\begin{equation}
p= \frac{1}{1+m},
\label{eq2}
\end{equation}

\noindent
where $p$ is the probability of the success in each trial and $m$ is the distribution mean.

%
\sidecaptionvpos{figure}{t}
\begin{SCfigure}
\centering
\includegraphics[width=0.5\textwidth]{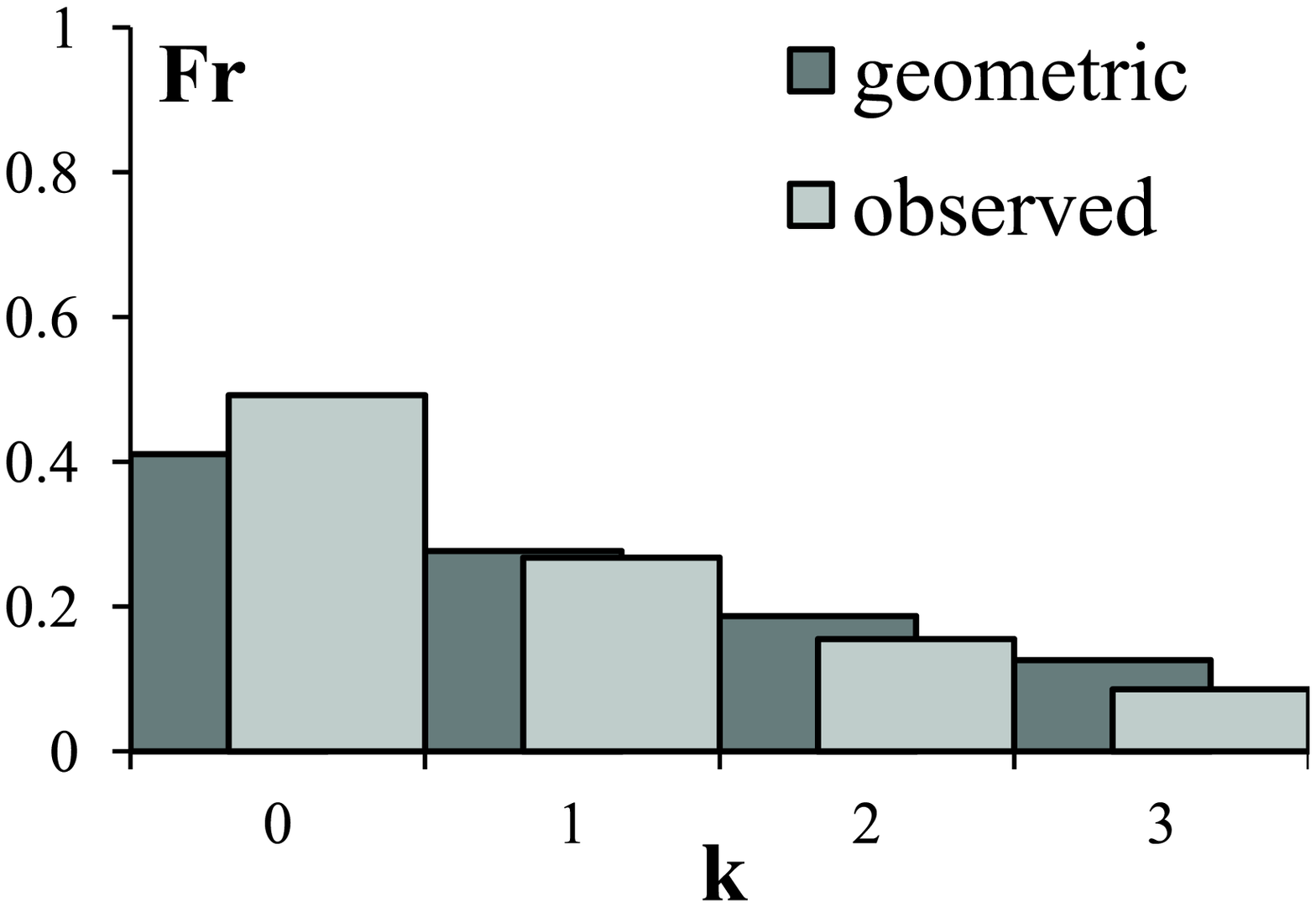}
\caption{Comparison of the observed $FD$ distribution and calculated geometric distribution ($FD$ relative frequencies, $F_r$ for different FD magnitude ranges, $k$) for the whole sample of 187 events.}
\label{fig3}
\end{SCfigure}

For our sample, the following association is made between the number of trials, $k$ and FD magnitude ranges:
\begin{itemize}
\item $k=0 \longleftrightarrow FD<1\,\%$;
\item $k=1 \longleftrightarrow 1\%<FD<3\,\%$;
\item $k=2 \longleftrightarrow 3\%<FD<6\,\%$;
\item $k=3 \longleftrightarrow FD>6\,\%$.
\end{itemize}
Using $k \leftrightarrow FD$ association we obtain the relative frequency distribution of $FD$ for our sample of 187 events (Figure \ref{fig3}). We calculate the distribution mean, $m=2.07$, and then using Equations (\ref{eq1}) and (\ref{eq2}) with renormalization (so that the total probability on all trials equals one) we construct the shifted geometric distribution for the whole sample of 187 events. In Figure \ref{fig3} we compare the distribution for our sample reconstructed using shifted geometric distribution to the observed distribution and a good agreement can be seen. A similar distribution of the observed FDs was obtained by \inlinecite{belov09}.

Since our sample in general follows the shifted geometric distribution, we assume that the shifted geometric distribution can describe the probability distribution of FD magnitude $[FD]$ for a \textit{certain} event with a specific set of remote solar observations of a CME and the associated flare. It was shown in Figures \ref{fig1} and \ref{fig2} that the trend of the change in the $FD$ distribution mean with a specific solar parameter can be fitted by a corresponding function. Therefore, based on the relationships found between $FD$ and solar parameters, a corresponding shifted geometric distribution can be obtained using Equations (\ref{eq1}) and (\ref{eq2}) for each solar parameter. We treat the empirical distribution obtained as a probability distribution for a specific solar parameter $\alpha$, where $\alpha=v,w,r,f,i$ (\textit{i.e.} initial CME speed $[v]$, CME apparent width $[w]$, CME/flare source position distance from the center of the solar disc $[r]$, flare strength $[f]$, and interaction parameter $[i]$). The probability distribution for a specific parameter provides the information on the probability for associating it with a specific value of $k$, \textit{i.e.} FD magnitude range. To combine the effect of solar parameters, they were treated as mutually non-exclusive and independent, in which case a joint probability distribution is given by \inlinecite{dumbovic15a}:


\begin{equation}
\begin{split}
P(FD=k) &= \sum_{\alpha}{P_{\alpha}}-\sum_{\alpha \ne \beta}{P_{\alpha}P_{\beta}}+\sum_{\alpha\ne \beta \ne \gamma}{P_{\alpha}P_{\beta}P_{\gamma}} -\\
 					 &-\sum_{\alpha \ne \beta \ne \gamma \ne \delta}{P_{\alpha}P_{\beta}P_{\gamma}P_{\delta}}+\sum_{\alpha \ne \beta \ne \gamma \ne \delta \ne \epsilon}{P_{\alpha}P_{\beta}P_{\gamma}P_{\delta}P_{\epsilon}}\,,
\end{split}
\label{eq3}
\end{equation}

\noindent
where $P_{\alpha}$=$P(\alpha)$ represents the probability of a specific FD magnitude range ($FD \leftrightarrow k$) for a specific solar parameter $\alpha$.

%
\begin{table}
\caption{CME/flare input parameters $[\alpha]$ and corresponding calculated geometric distribution parameters $[m(\alpha)$ and $p(\alpha)]$ for two extreme events}
\label{tab2}
\begin{tabular}{cccccc}
\cline{1-6}
\multicolumn{3}{c}{EVENT 1} & \multicolumn{3}{c}{EVENT 2}\\
\cline{1-6}
$\alpha$ & $m(\alpha)$ & $p(\alpha)$ & $\alpha$ & $m(\alpha)$ & $p(\alpha)$\\
\cline{1-6}
$v=2000$ km\,s\,$^{-1}$           & 3.79 & 0.2089 & $v=450$ km\,s\,$^{-1}$          & 0.69 & 0.5924\\
$w=360^{\circ}$                   & 3.81 & 0.2079 & $w=50^{\circ}$                  & 0.71 & 0.5848\\
$r=0.05$ R$_{\odot}$              & 3.76 & 0.2103 & $r=0.99$ R$_{\odot}$            & 1.03 & 0.4929\\
$f=5000$x$10^{-7}$ W\,m\,$^{^-2}$ & 4.11 & 0.1957 & $f=10$x$10^{-7}$ W\,m\,$^{^-2}$ & 0.38 & 0.7238\\
$i=4$                             & 2.70 & 0.2703 & $i=1$                           & 1.20 & 0.4545\\
\cline{1-6}
\end{tabular}
\end{table}

Using the found relationships between $FD$ and solar parameters from Figures \ref{fig1} and \ref{fig2}, and Equations (\ref{eq1})\,--\,(\ref{eq3}) we calculate a probability distribution for two extreme events: EVENT 1, which was a very fast and wide CME, involved in a CME--CME interaction and associated with a strong X-class flare close to the center of the solar disc (presumably intensly GCR-effective), and EVENT 2, which was a slow and narrow CME, which was not involved in a CME--CME interaction and is associated with a weak B-class flare near the limb of the solar disc  (presumably not GCR-effective). The input CME/flare parameters for both of these extreme events is given in Columns 1 and 3 in Table \ref{tab2}, respectively. Using the relationships between $FD$ and solar parameters from Figures \ref{fig1} and \ref{fig2} (suitable for the given units), we obtain the distribution mean for each of the solar parameters $[m(\alpha)]$ (Columns 2 and 4 in Table \ref{tab2}). It can be seen that $m(\alpha)$ attains smaller values for EVENT 2, as expected (the distribution is shifted towards smaller FD magnitudes). Using Equation (\ref{eq2}), we obtain the corresponding probability of success in each trial for each of the solar parameters, $[p(\alpha)]$ (Columns 3 and 6 in Table \ref{tab2}), where the shift of the FD distribution for the EVENT 2 towards smaller FD magnitudes is reflected by the increased values of $p(\alpha)$.

\begin{figure}
\centering
\includegraphics[width=0.95\textwidth]{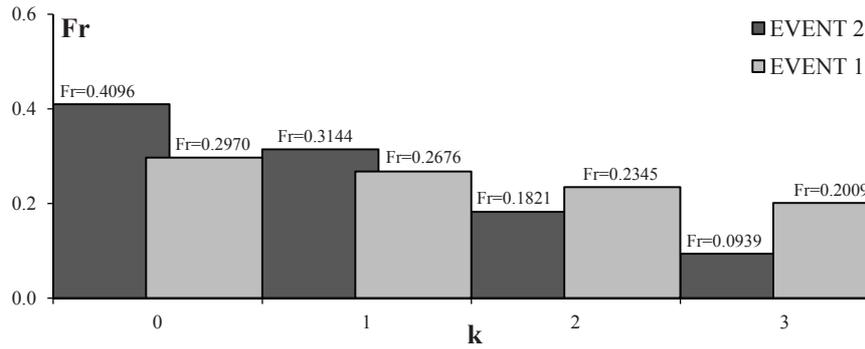}
\caption{The joint probability distribution for EVENT 1 (light grey) and EVENT 2 (dark grey). The values of relative frequencies are given above the corresponding $k$-bin.}
\label{fig4}
\end{figure}

Using Equation (\ref{eq1}) the relative frequency for each trial [$k=0,1,2,3$] and each solar parameter [$\alpha=v,w,r,f,i$] can be calculated for each of the two extreme events. Finally, using Equation (\ref{eq3}), we calculte the joint probability distribution, \textit{i.e.} the relative frequency for a given set of solar parameters [$v,w,r,f,i$] for each trial $[k]$ and renormalize it so that the total probability equals one ($\sum_{k=0}^{3}{P(k)}=1$). The resulting distribution represents the joint probability distribution of observing FD magnitude $[FD]$ in a specific range [$FD<1\,\%$, $1\,\%<FD<3\,\%$, $3\,\%<FD<6\,\%$, $FD>6\,\% \leftrightarrow k=0$, $k=1$, $k=2$, $k=3$] for a CME/flare event with a specific set of solar parameters [$v$, $w$, $r$, $f$, $i$]. The joint probability distributions for EVENT 1 and EVENT 2 are shown in Figure \ref{fig4}.

%

It can be seen in Figure \ref{fig4} that the distribution for the two extreme events is different and that the probability for higher FD magnitudes is larger for EVENT 1, representing faster and wider CMEs that originate near the disc center, are related to more energetic flares and are likely to be involved in a CME–-CME interaction. However, in both distributions the highest probability is that the event will not be GCR-effective, \textit{i.e.} that FD magnitude will be $FD<1\,\%$ ($k=0$). Although the probability distribution changes with CME/flare parameters it is always highly asymmetric with the greatest probability that CME will not be GCR-effective. This depicts the general behavior of CMEs seen in Figure \ref{fig3}: a large majority of CMEs will never reach the Earth and/or will not be very GCR-effective. Therefore, the probability distribution does not give a straightforward prediction of whether or not (and how large) Forbush decrease will be. Therefore, the level of GCR-effectiveness needs to be obtained by imposing some criteria (thresholds) on the probability distribution.

%
\begin{figure}
\centerline{\includegraphics[width=0.5\textwidth]{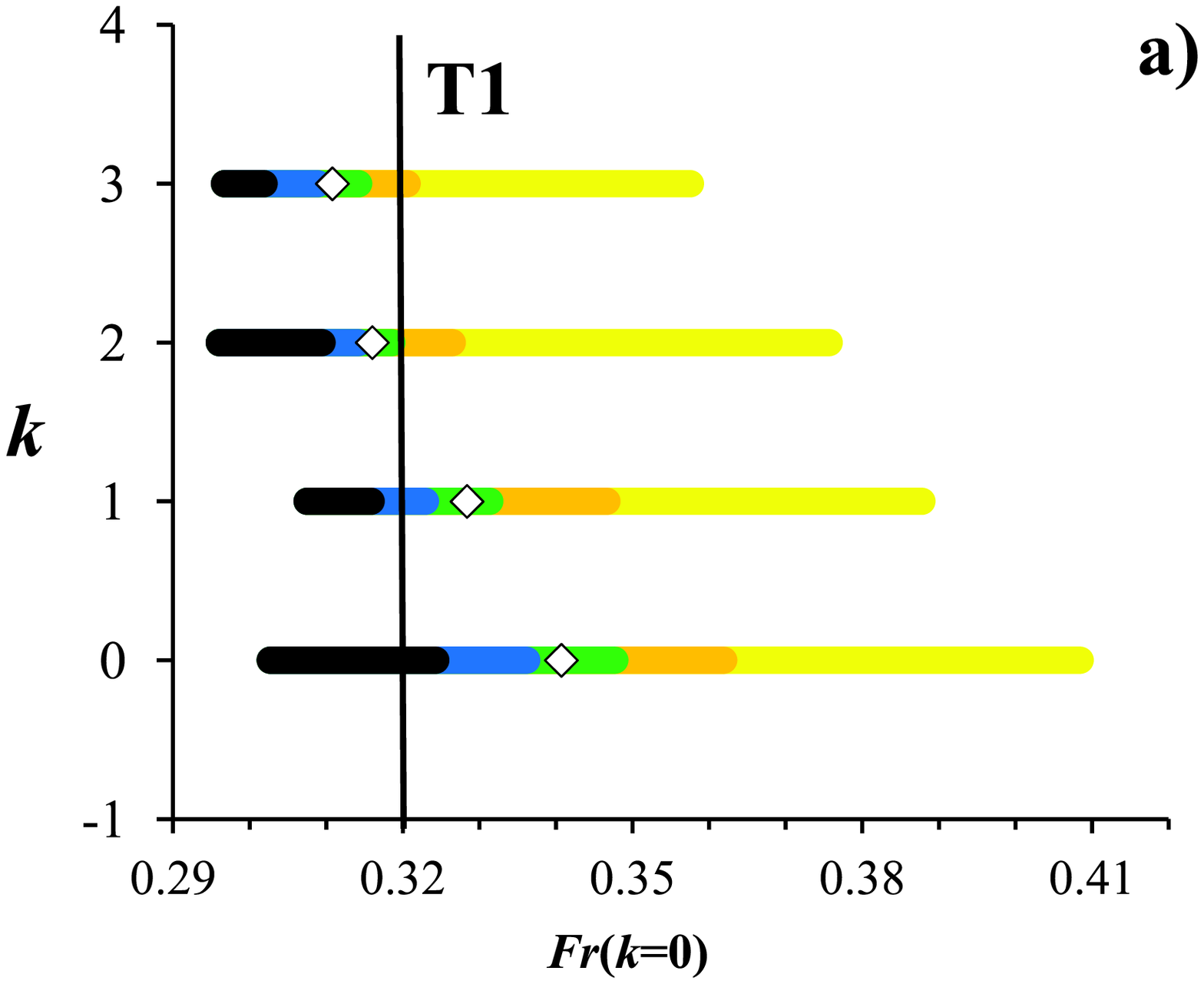}
            \hspace*{0.03\textwidth}
            \includegraphics[width=0.5\textwidth]{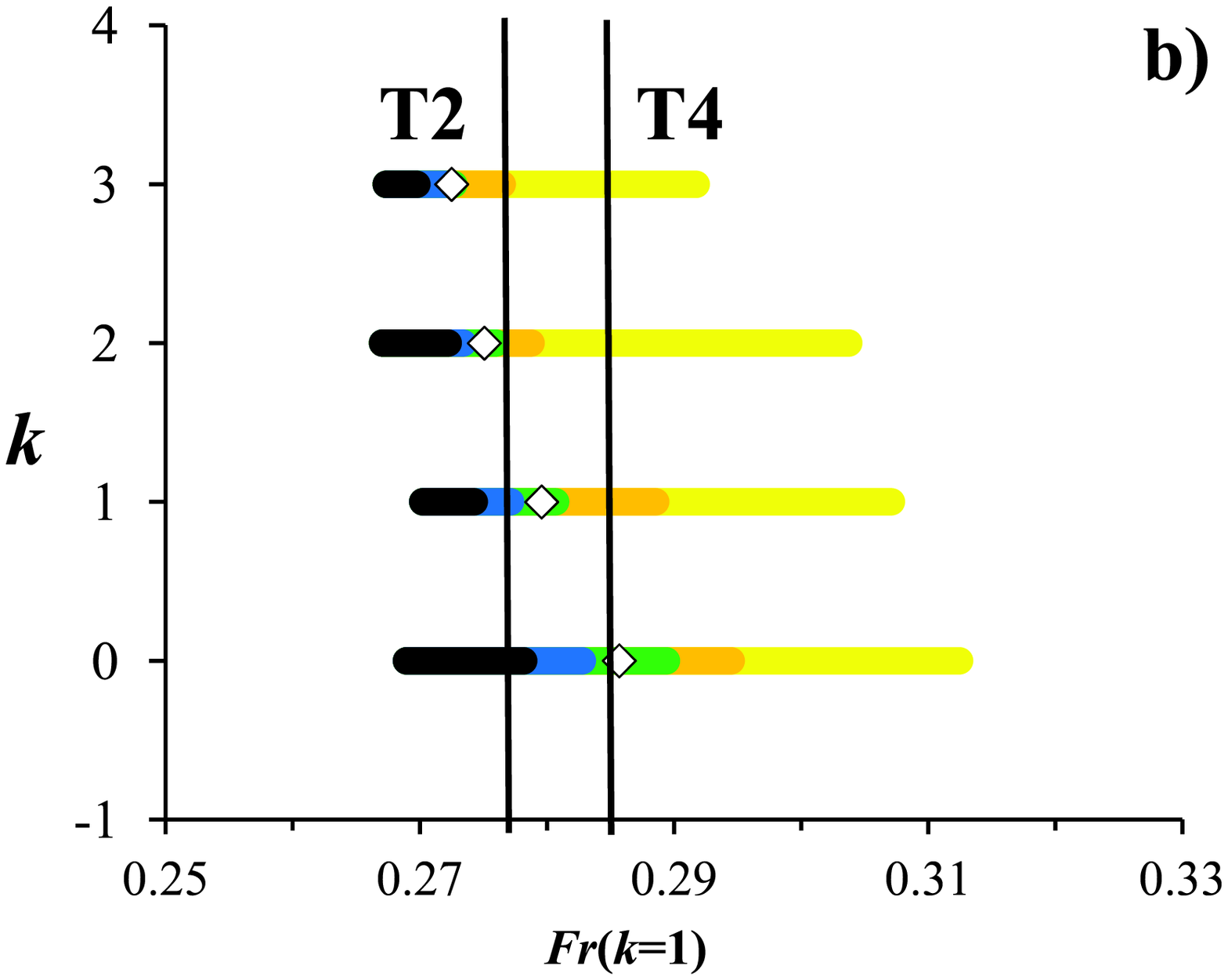}
              }
\vspace{0.03\textwidth}
\centerline{\includegraphics[width=0.5\textwidth]{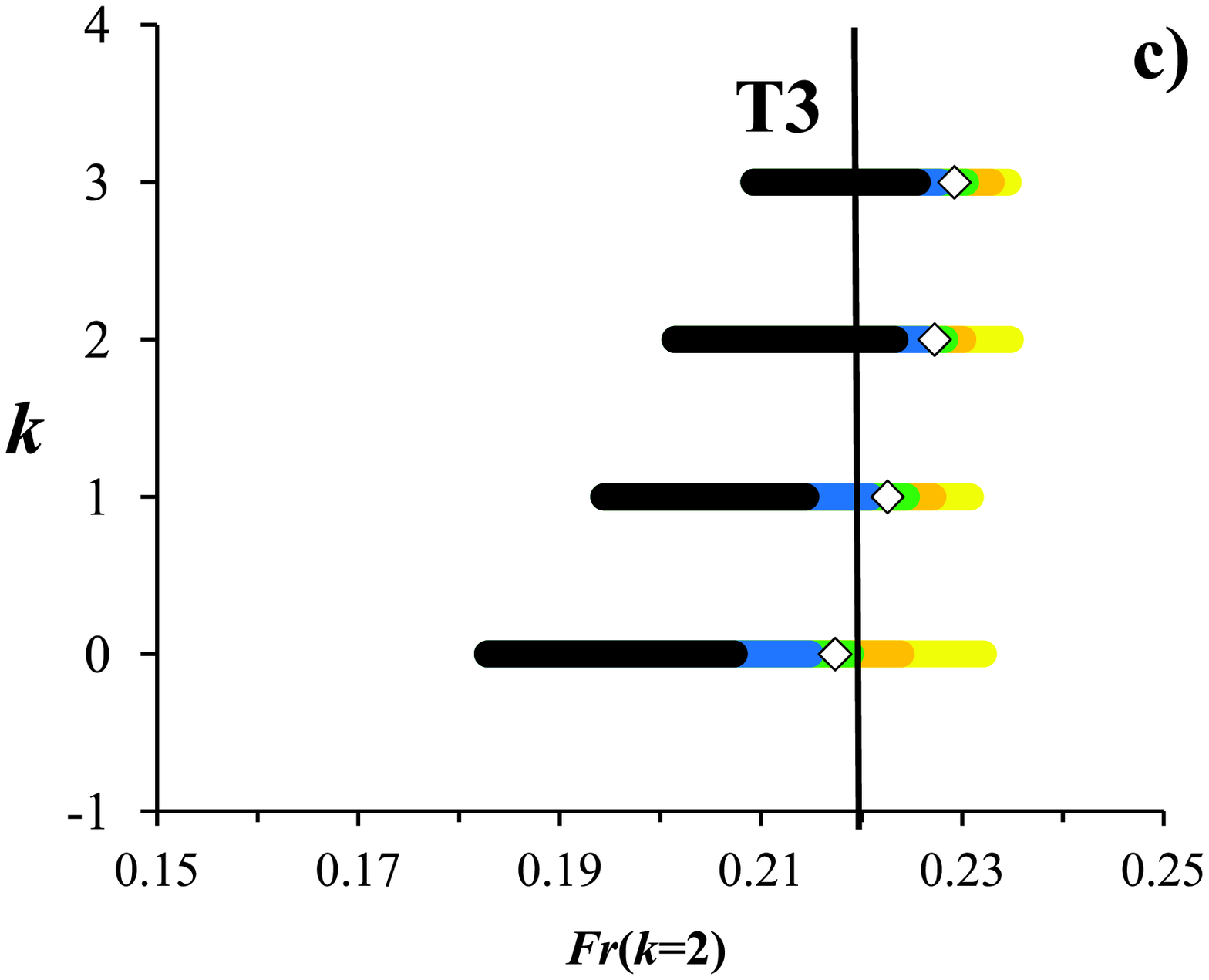}
            \hspace*{0.03\textwidth}
            \includegraphics[width=0.5\textwidth]{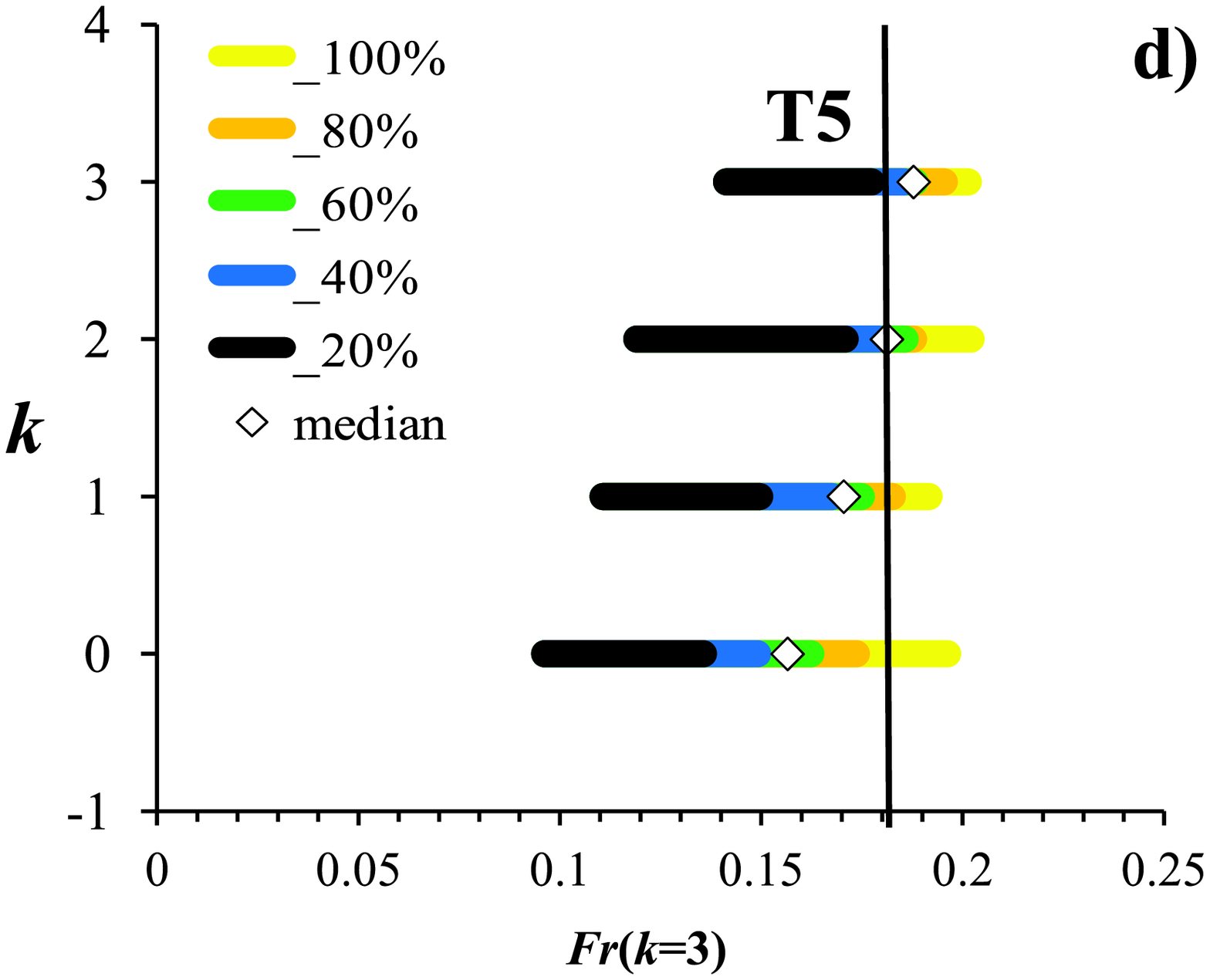}
              }
\caption{Density plots representing data scatter of the calculated relative frequencies $[F_r(k)]$ against the observed FD magnitude ranges $[k]$ for 187 CME--flare--Dst--FD associations. The density of the data points is expressed by differently colored percentiles. White diamond marks median, whereas black lines mark established thresholds T1\,--\,T5 (for explanation see main text).}
\label{fig5}
\end{figure}

In Figure \ref{fig4} we can see that for low GCR-effectiveness (EVENT 2) we expect a much higher value of relative frequency for $k=0$ $[F_r(k=0)]$  than for the highly GCR-effective event (EVENT 1). Conversely, we expect a much higher value of relative frequency for $k=3$ $[F_r(k=3)]$ for a highly GCR-effective event (EVENT 1), than for a low GCR-effectiveness (EVENT 2). Therefore, thresholds on the value of the relative frequency for a certain bin can be established to enclose a certain range of GCR-effectiveness. These thresholds are derived empirically. For that purpose we use the list of 187 CME--flare--Dst--FD associations and calculate FD magnitude distribution  for each of the events in the list, based on the corresponding CME/flare parameters. Therefore, for each event we obtain four different relative frequency values $[F_r(k)]$ corresponding to four different distribution bins: $k=0,1,2,3$. For each relative frequency $[F_r(k)]$ we produce a scatter plot against the observed $FD$ value, where $FD$ is expressed as one of the four possible FD magnitude ranges associated with four different $k$: [$FD<1\,\%$, $1\,\%<FD<3\,\%$, $3\,\%<FD<6\,\%$, $FD>6\,\% \leftrightarrow k=0$, $k=1$, $k=2$, $k=3$]. Since $FD$ is given by four discrete values, the data in these plots will be scattered in four lines at $k=0,1,2,3$. Each of the lines contains a number of data points that corresponds to the number of observations of different FD magnitude range (92 events with $k=0$, 50 events with $k=1$, 29 events with $k=2$, and 16 events with $k=3$). We find it useful to present the scatter plot for each of these four lines as a density plot using percentiles. In that way, it is noticeable how many data points are encompassed in each $F_r(k)$. Using the density of data scatter as a guideline, we derive thresholds T1\,--\,T5 as values which best separate different GCR-effectiveness. These density plots representing the data scatter of the calculated relative requencies $[F_r(k)]$ against the observed FD magnitude, as well as thresholds T1\,--\,T5, are presented in Figure \ref{fig5}.

It can be seen in Figure \ref{fig5}a that almost 80\,\% of $k=3$ events and 60\,\% of $k=2$ events have $F_r(k=0)<0.32\equiv T1$, whereas more than 80\,\% of non GCR-effective events ($k=0$) and more than 60\,\% of moderately GCR-effective events ($k=1$) have $F_r(k=0)>T1$. Therefore, we establish $T1$ as a threshold separating $k=0,1$ events from $k=2,3$ events. Similarly, we obtain thresholds $T2$ and $T3$ in Figures \ref{fig5}b and \ref{fig5}c, respectively. Establishing a threshold between $k=0$ and $k=1$ events is more challenging, since the difference in the data density is less pronounced compared to separating $k=0,1$ and $k=2,3$ events. In Figure \ref{fig5}b a threshold $T4$ is shown, which separates $k=0$ data (more than 50\,\% of events have $F_r(k=1)>T4$) from $k=1$ data (more than 60\,\% of events have $F_r(k=1)<T4$). The difference in the data density is even less pronounced in separating $k=2$ and $k=3$ events. In Figure \ref{fig5}d a threshold $T5$ is shown, which separates $k=2$ data (more than 50\,\% of events have $F_r(k=3)<T5$) from $k=3$ data (more than 50\,\% of events have $F_r(k=3)>T5$).

%
\sidecaptionvpos{figure}{t}
\begin{SCfigure}
\centering
\includegraphics[width=0.6\textwidth]{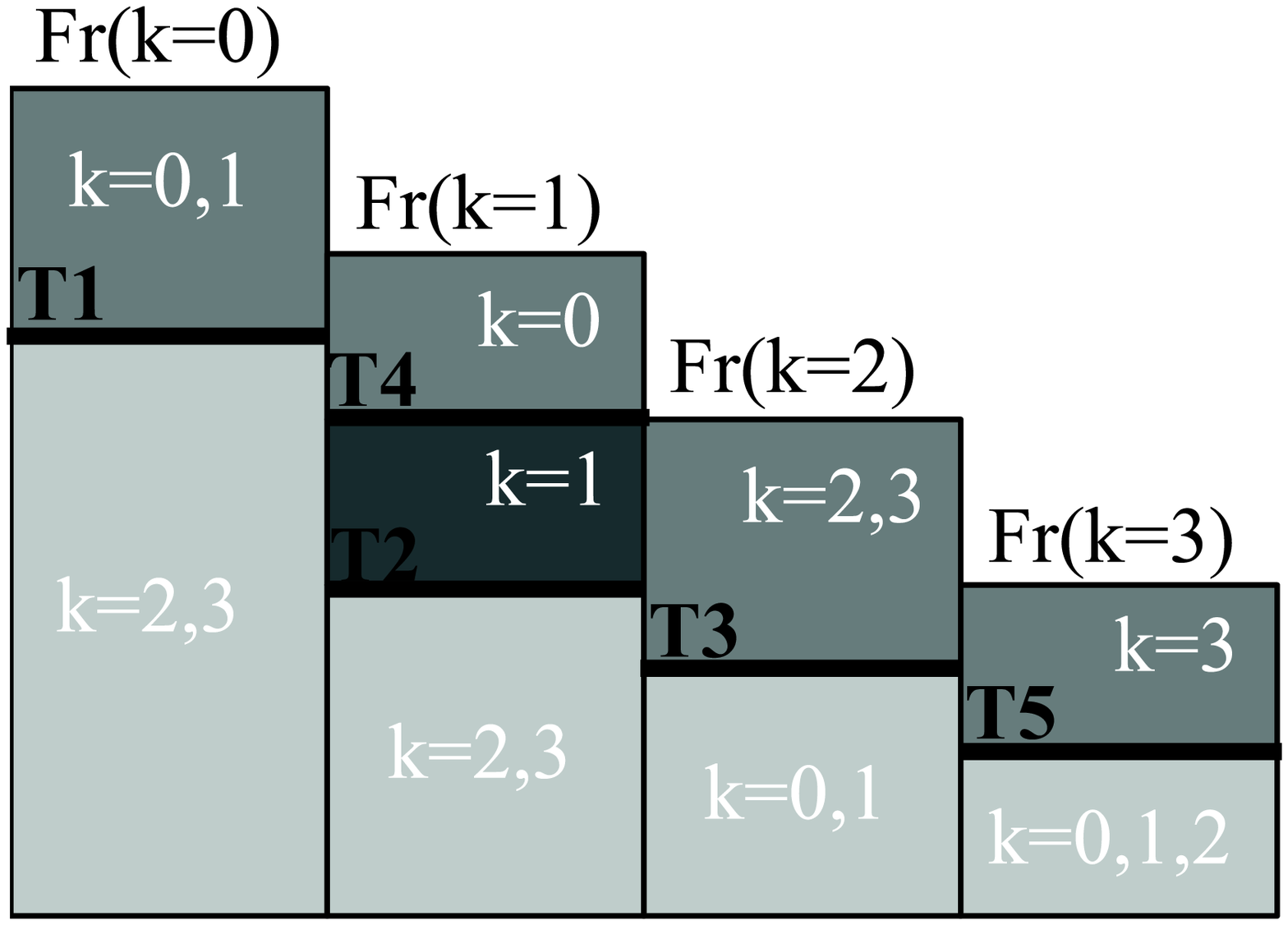}
\caption{Schematic of thresholds for relative frequencies of certain bins $F_r$$(k)$: $T_1=0.32$, $T_2=0.277$, $T_3=0.222$, $T_4=0.285$, and $T_5=0.183$. Possible GCR-effectiveness level $k$ is given for values above/below the corresponding threshold.}
\label{fig6}
\end{SCfigure}

We interpret the thresholds as values that encompass most of the events with a certain GCR-effectiveness. For example, most of the intense GCR-effective events ($FD>6\,\% \leftrightarrow k=3$)  have a relative frequency for $k=3$, $F_r(k=3)>T5$. Therefore, we expect that if $F_r(k=3)<T5$ the event will be intensly GCR-effective ($FD>6\,\% \leftrightarrow k=3$), otherwise it will be less GCR-effective, \textit{i.e.} it will have some other level of GCR-effectiveness: \{$FD<1\,\%$, $1\,\%<FD<3\,\%$, $3\,\%<FD<6\,\%$\} $\leftrightarrow$ \{$k=0$, $k=1$, $k=2$\}. A schematic of the thresholds for relative frequencies of certain bins is given in Figure \ref{fig6}. Conditions for some of the thresholds immediately give the information on which GCR-effectiveness level is expected. However, for some thresholds there still remains a set of possible GCR-effectiveness levels. Combining conditions for different thresholds, a unique GCR-effectiveness level can be obtained. The conditions for determing the GCR-effectiveness level using thresholds are given in Table \ref{tab3}.

For example, when we apply the first three conditions from Table \ref{tab3} to the joint probability distribution for EVENT 1 (Figure \ref{fig4}) we derive the following: $F_r$$(k=3)>T_1$, $F_r$$(k=2)>T_2$, and $F_r$$(k=0)<T_3$. All three conditions are in favor of $k=2,3$; therefore we apply the final condition from Table \ref{tab3} and find that $F_r$$(k=1)<T_5$, which means that the expected FD magnitude is $k=3 \leftrightarrow FD>6\,\%$. We repeat the calculation for EVENT 2 from Figure \ref{fig4}, where the first three conditions from Table \ref{tab3} result in $F_r$$(k=3)<T_1$, $F_r$$(k=2)<T_2$, and $F_r$$(k=0)>T_3$ being in favor of $k=0,1$. We then apply the fourth condition from Table \ref{tab3} and find that $F_r$$(k=1)>T_4$, which means that the expected FD magnitude is $k=0 \leftrightarrow FD<1\,\%$. Therefore, starting from extremely different solar CME/flare parameters we derive two extremes of GCR-effectiveness level (EVENT 1 is intensly GCR-effective, whereas EVENT 2 is not GCR-effective).

%
\begin{table}
\caption{The conditions for determing the GCR-effectiveness level using thresholds [$T_i$, $i=1,2,3,4,5$] for relative frequencies of certain bins $[F_r$$(k)]$ given in Figure \ref{fig5}. Combination of these conditions give a unique GCR-effectiveness level.}
\label{tab3}
\begin{tabular}{cccc}     
\hline
condition based & result & result & description \\
on thresholds & (if satisfied) & (if not satisfied) & of the conditions\\
\hline
$F_r$$(k=3)<T_1$ & $k=0,1$ & $k=2,3$ & the combination of the first\\
$F_r$$(k=2)<T_2$ & $k=0,1$ & $k=2,3$ & three conditions determines\\
$F_r$$(k=0)>T_3$ & $k=0,1$ & $k=2,3$ & whether $k=0,1$ or $k=2,3$\\
\hline
                 &         &         & once established that $k=0,1$\\
$F_r$$(k=1)>T_4$ & $k=0$   & $k=1$   & this condition determines\\
                 &         &         & whether $k=0$ or $k=1$\\
\hline
                 &         &         & once established that $k=2,3$\\
$F_r$$(k=1)>T_5$ & $k=2$   & $k=3$   & this condition determines\\
                 &         &         & whether $k=2$ or $k=3$\\
\hline
\end{tabular}
\end{table}

The model is empirical and based on the remote solar CME/flare observations of the sample used; therefore, the model input has certain limitations. CME speed $[v]$ is a continuous parameter given in units km\,s\,$ ^{-1}$ in the range $v>106$, restricted by the $v$-intercept in Figure \ref{fig2}a. The CME/flare source distance from the center of the solar disc $[r]$ is also a continuous parameter given in units of solar radii, with the range restricted by the physical boundaries, $0\le r\le1$ (\textit{i.e.} the center of the solar disc and the solar limb). The apparent width $[w]$ is a continuous parameter restricted to the range $0^{\circ}<w\le360^{\circ}$, determined by observational boundaries ($w=0^{\circ}$ means a CME was not detected, $w=360^{\circ}$ is a halo CME). The flare strength parameter $[f]$, \textit{i.e.} flare soft X-ray peak intensity, is a continuous parameter $[10^{-7}$W\,m\,$^{ ^-2}]$ in the range $f>5.3$ restricted by the $f$-intercept in Figure \ref{fig2}c. Finally, the interaction parameter $[i]$ is  a discrete parameter that can attain values $i=1,2,3,4$ based on the likeliness of the CME--CME interaction (Section \ref{data}).

Finally, we consider the implications of the approximation, where the solar parameters were treated as independent and Equation (\ref{eq3}) applies. As noted by \inlinecite{dumbovic15a} this assumption is not fully valid, due to the fact that not all key solar parameters are independent of each other and this assumption may increase to some extent the number of false alarms. However, we note that this assumption substantially simplifies the calculation of the joint probability, while on the other hand, the number of the false alarms is greatly influenced by the fact that the probability distribution is highly asymmetrical and in favor of low GCR-effectiveness, and even the optimized thresholds do not enclose \textit{all} of the events (only a majority of events).

\section{Evaluation of the Prediction}
\label{evaluation}

The prediction was evaluated first using the \textbf{training set}, \textit{i.e.} the sample of the 187 CME--flare--FD associations used for the statistical analysis. The evaluation applied to the training set describes the successfulness and the reliability of the prediction model with respect to the approximations used, since we assume that our sample represents the ensemble of possibilities for a certain event. Next we perform the evaluation using a \textbf{test set}, \textit{i.e.} independent sample of additionally selected and measured 42 CME--flare--FD events. We note that the two sets are conveniently named in analogy with the neural network approach (see, \textit{e.g.}, \opencite{valach09}; \opencite{uwamahoro12}; \opencite{sudar15}), which typically uses three different sets (training set, validating set, and test set), with the difference that in our case the validating set is identical to the training set. The test set comprises of events in the time period 1998\,--\,2012, which are not present in the training set. The method for FD association is the same as for the training set (described in Section \ref{data}), with the difference that the CR data after 2011 were taken from the Neutron Monitor Database event search tool (\url{www.nmdb.eu/nest/search.php}) from Kiel, Magadan, and Newkirk stations.

%
\begin{table}
\caption{Contingency table for a binary event}
\label{tab4}
\begin{tabular}{|c|c|c|c|}     
\cline{3-4}
\multicolumn{2}{c|}{} & \multicolumn{2}{c|}{Observation}\\
\cline{3-4}
\multicolumn{2}{c|}{} & YES & NO\\
\cline{1-4}
 & & a = number of hits, & b = number of false alarms,\\
 & YES & \textit{i.e.} correctly & \textit{i.e.} forecasts of an event while\\
Forecast & & forecasted events & no event was observed\\
\cline{2-4}
 & & c = number of misses,& d = number of correct rejections,\\
 & NO & \textit{i.e.} events which & \textit{i.e.} events which were not forecasted\\
 & & were not forecasted & while indeed no event was observed\\ 
\cline{1-4}
\end{tabular}
\end{table}

We evaluate the forecast by comparing the predicted value with the observed value using verification measures for binary events (see, \textit{e.g.}, \opencite{devos14}). The verification measures are defined based on the contingency table (Table \ref{tab4}), which describes four possible outcomes (hit, false alarm, miss, and correct rejection). For the purpose of the evaluation we redefine the ``event" as association of FD with a particular value. For example we define $k=0$ ($FD<1$\,\%) as an event. The event is classified as a ``hit" when $k=0$ was both observed and predicted; ``false alarm" is when $k=0$ is observed, while $k\ne0$ was predicted; ``miss" is when $k\ne0$ was observed, while $k=0$ was predicted; ``correct rejection" is when $k\ne0$ was both observed and predicted. 

We use the following verification measures (for more details see \opencite{devos14} and references therein):

\noindent \textit{The Probability Of Detection} (POD) or hit rate, the ratio of the number of hits and the number of events, calculated as POD$=a/(a+c)$;\\
\textit{The False Alarm Ratio} (FAR), the ratio of the number of false alarms and the total number of forecasts, calculated as FAR=$b/(a+b)$;\\
\textit{Bias} (BIAS), the ratio of the number of forecasts of occurrence to the number of actual occurrences, calculated as BIAS=$(a+b)/(a+c)$;\\
\textit{Heidke Skill Score} (HSS), skill score taking into account the number of correct random forecasts, calculated as HSS=$(a+d-E)/(n-E)$,\\
where $E=((a+c)(a+b)+(c+d)(b+d))/n$ and $n=a+b+c+d$

\begin{figure}
\centerline{\includegraphics[width=0.5\textwidth]{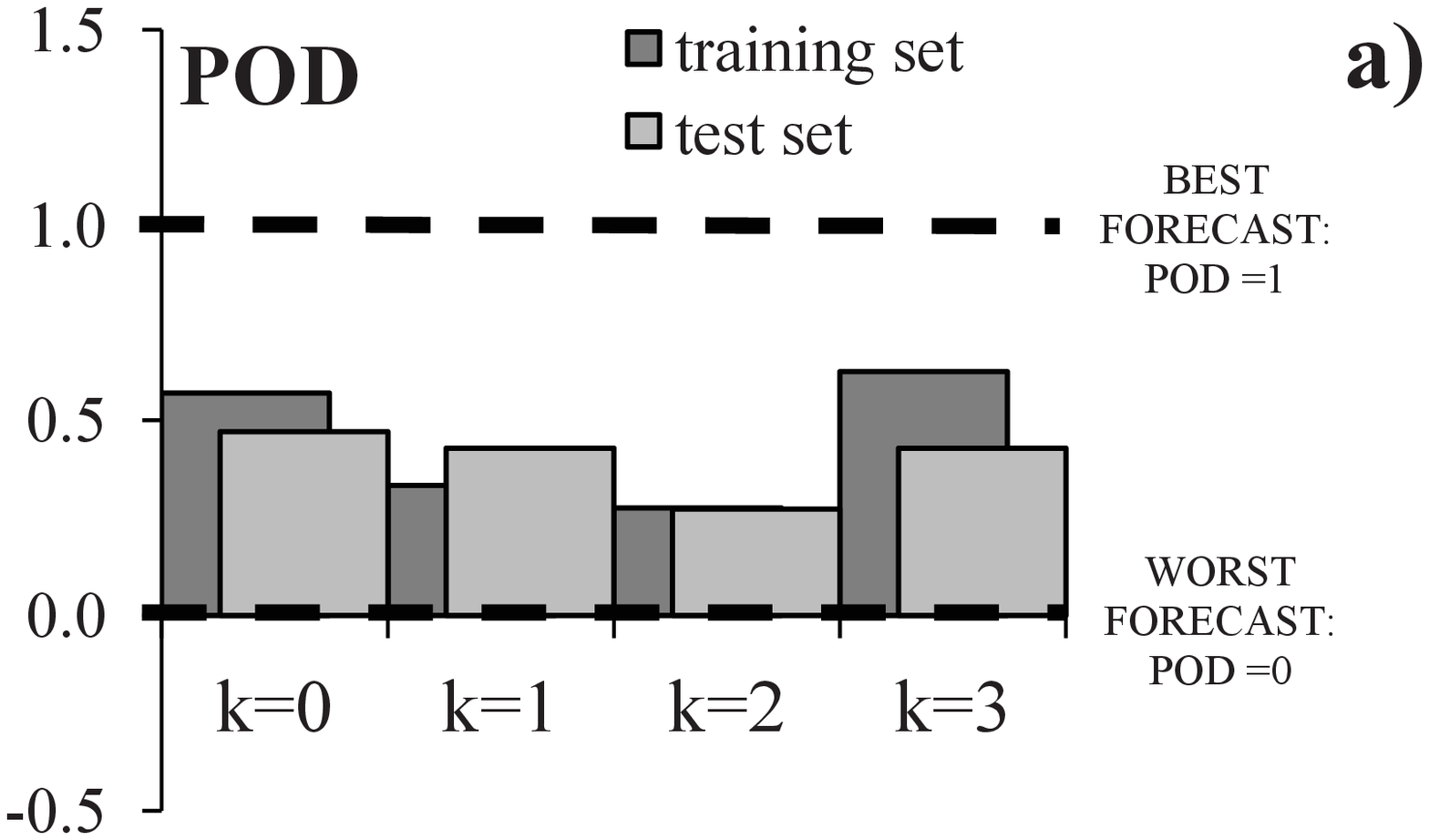}
            \hspace*{0.03\textwidth}
            \includegraphics[width=0.5\textwidth]{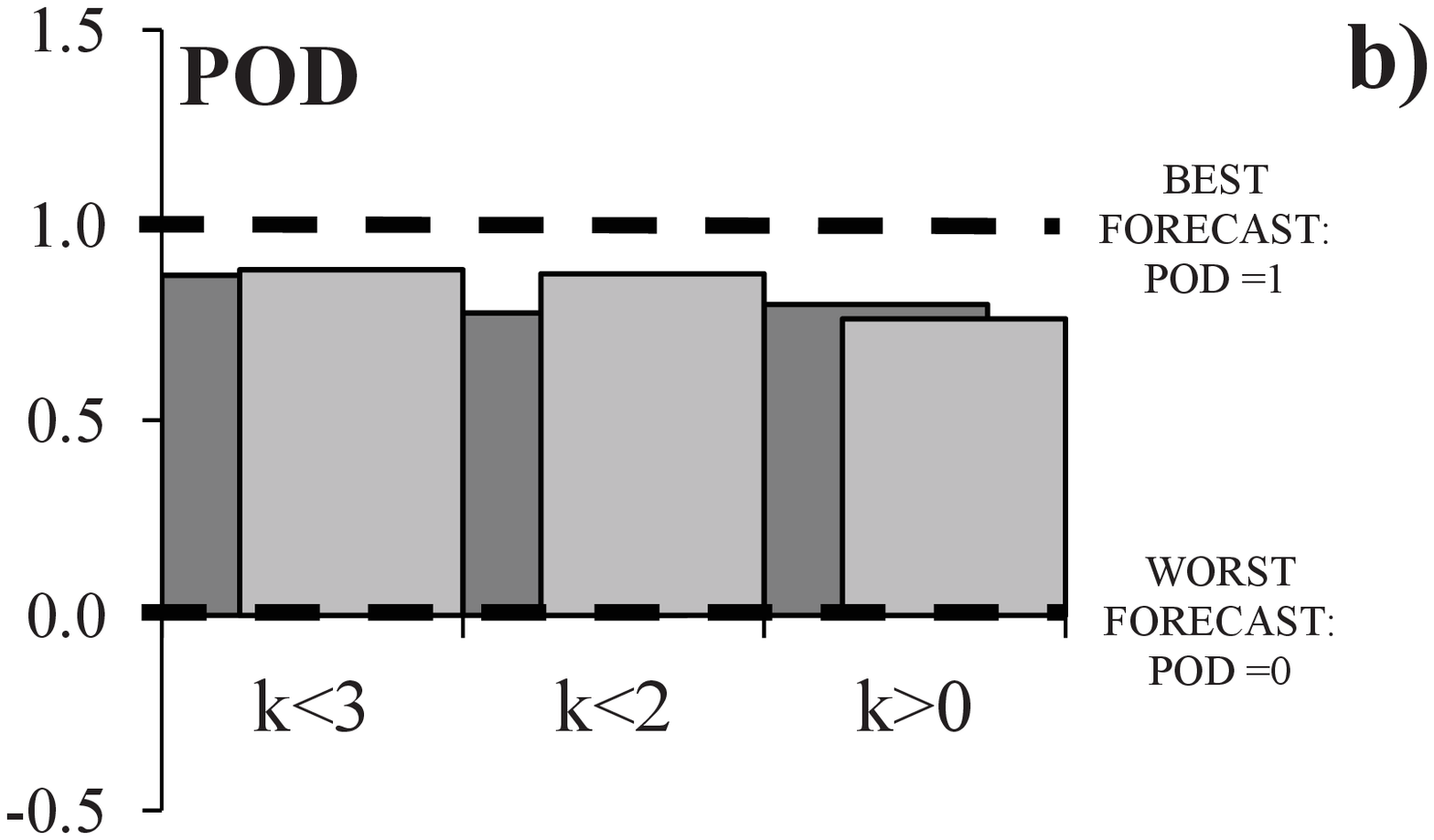}
              }
\vspace{0.03\textwidth}
\centerline{\includegraphics[width=0.5\textwidth]{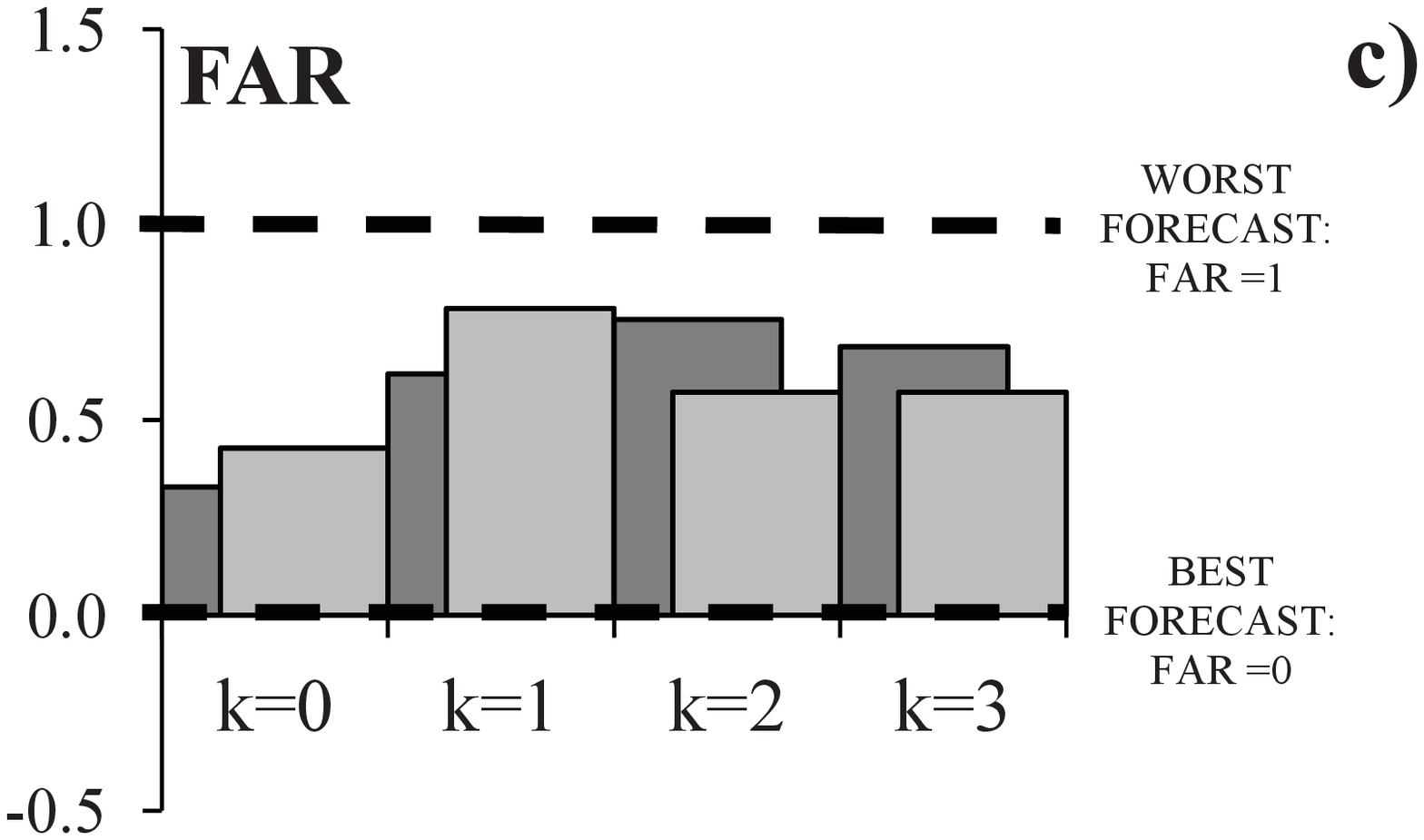}
            \hspace*{0.03\textwidth}
            \includegraphics[width=0.5\textwidth]{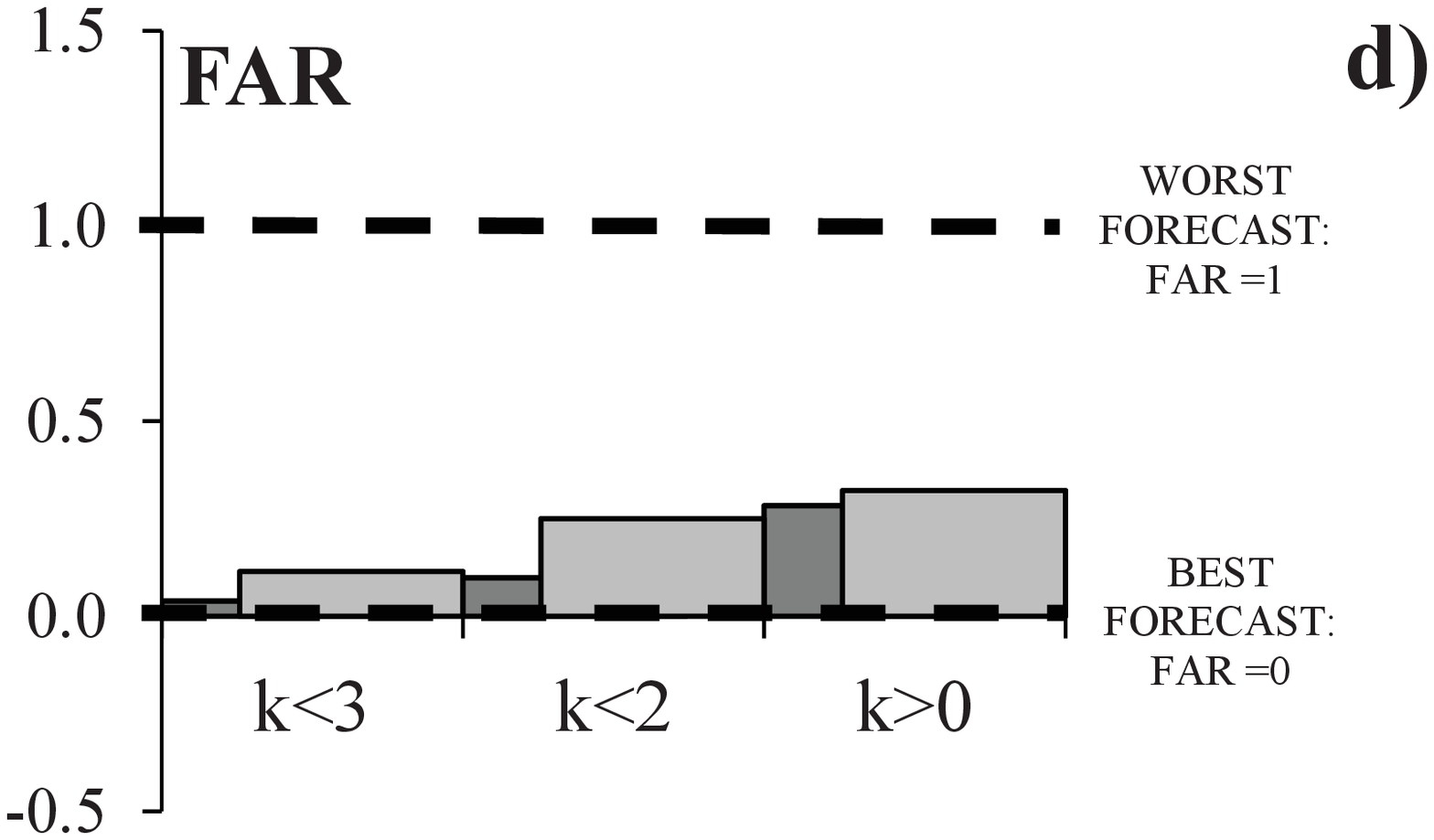}
              }
\vspace{0.03\textwidth}
\centerline{\includegraphics[width=0.5\textwidth]{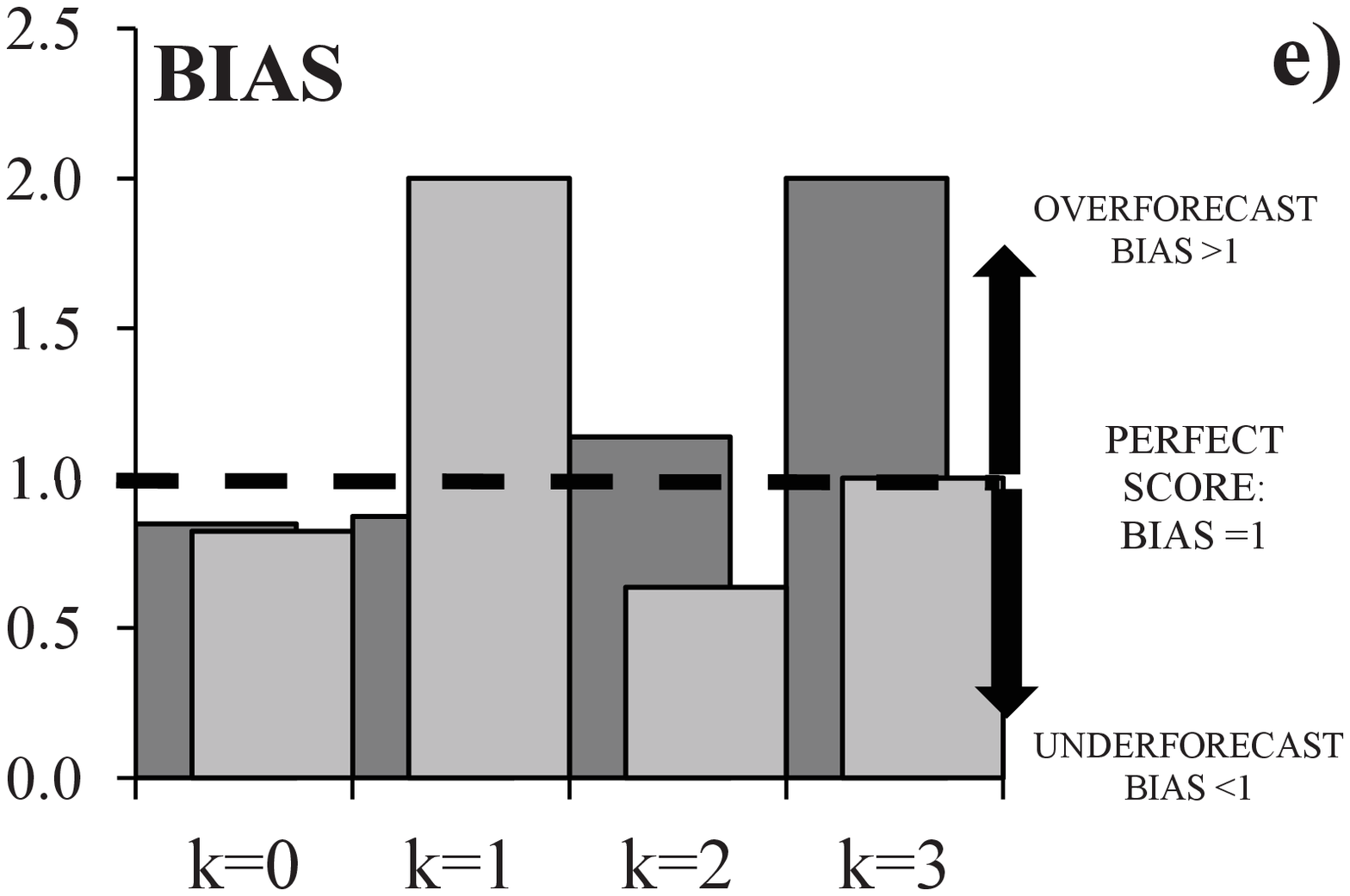}
            \hspace*{0.03\textwidth}
            \includegraphics[width=0.5\textwidth]{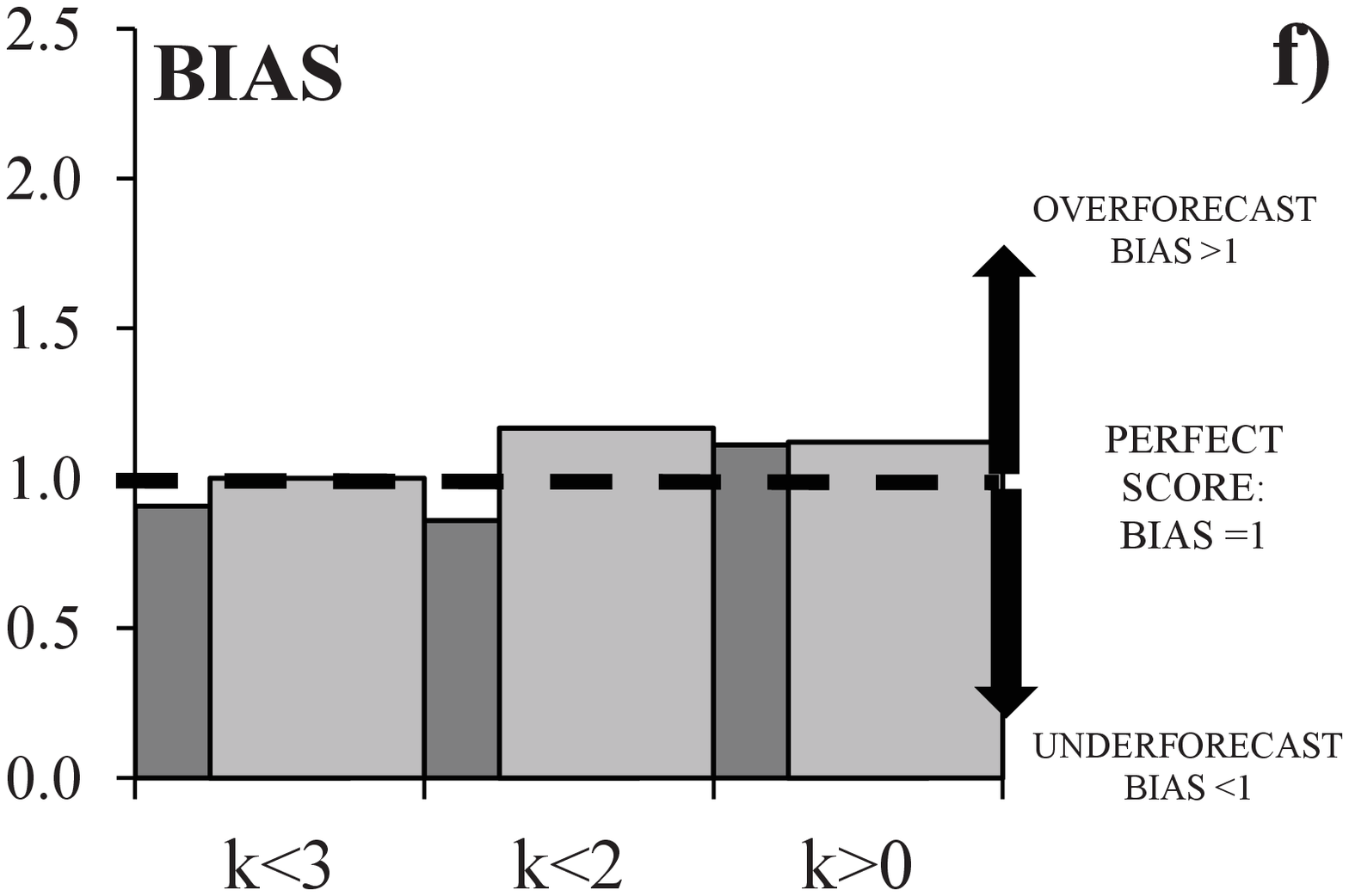}
              }
\vspace{0.03\textwidth}
\centerline{\includegraphics[width=0.5\textwidth]{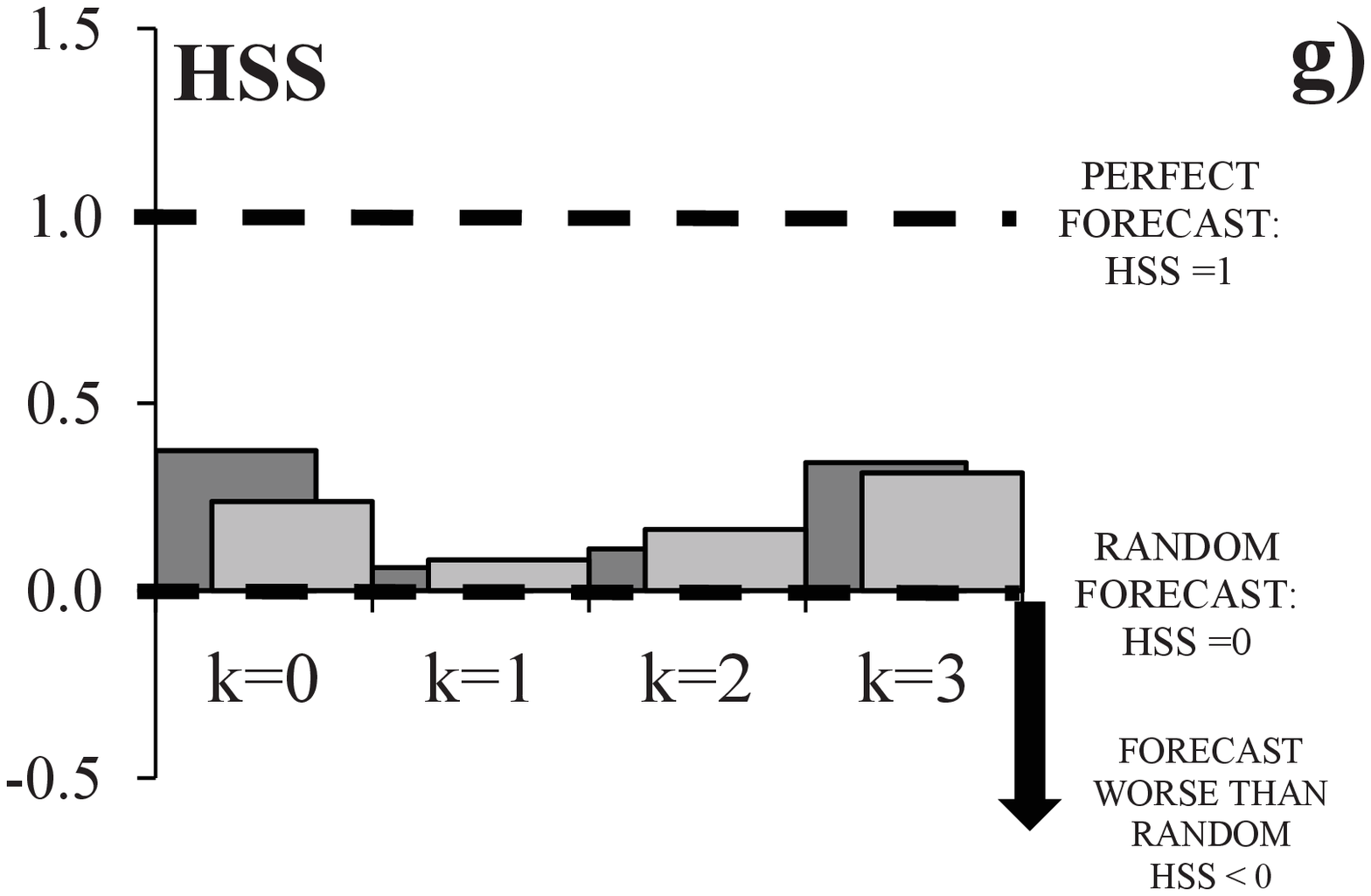}
            \hspace*{0.03\textwidth}
            \includegraphics[width=0.5\textwidth]{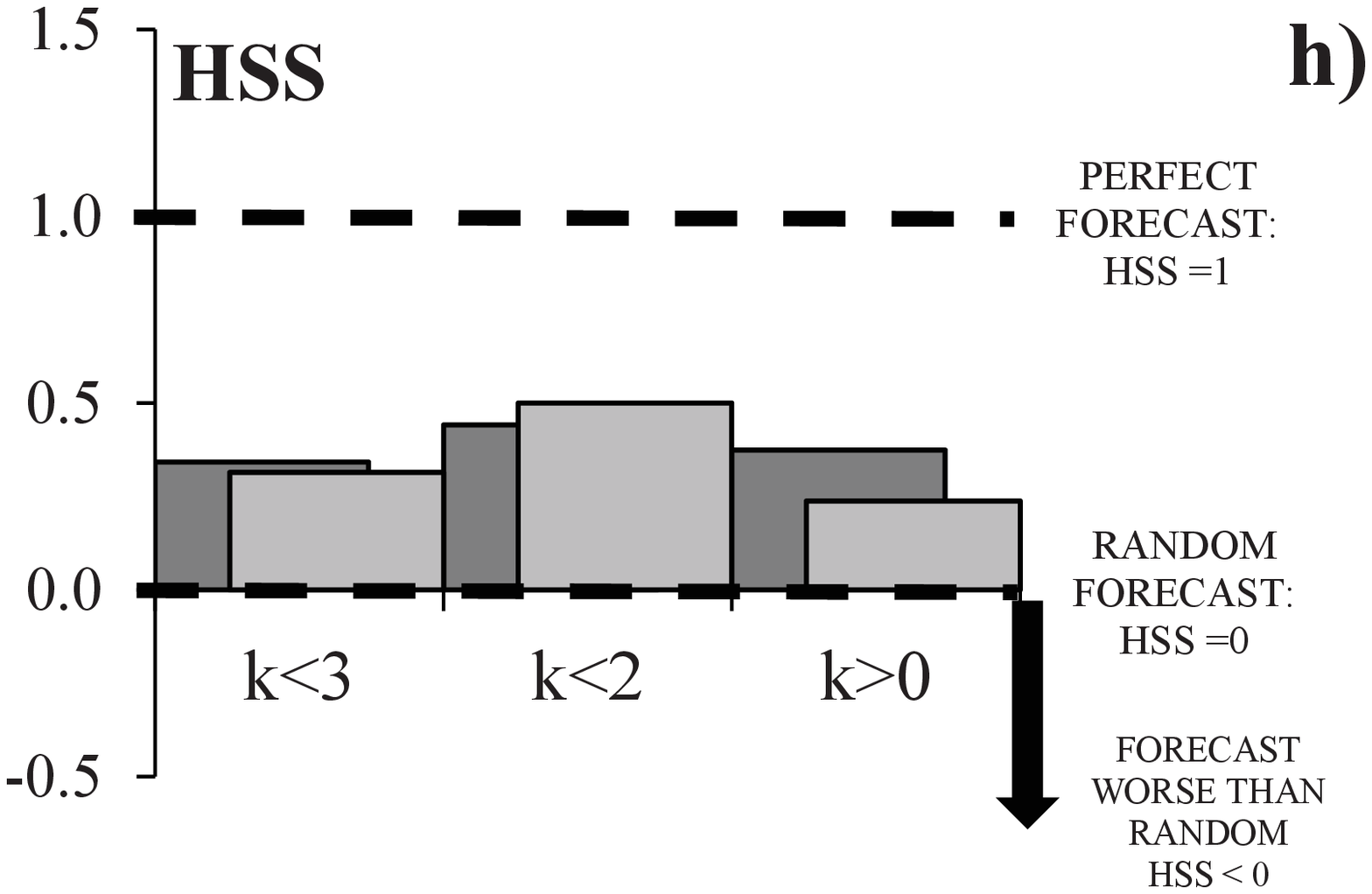}
              }                           
\caption{The Probability Of Detection (POD), False Alarm Ratio (FAR), BIAS, and Heidke Skill Score (HSS) for training and test sets for a more specific ($[k=0,1,2,3]$) and less specific ([$k<3$, $k<2$, and $k>0$]) forecast.}
\label{fig7}
\end{figure}

Each of the verification measures gives information on the quality of the prediction, however, none of them gives a full information on the quality of the forecast system. POD describes what fraction of the observed ``yes" events were correctly forecast and ranges from 0 to 1, with perfect score POD\,=\,1 (all hits). It is sensitive to hits, but ignores false alarms and therefore should be used in conjunction with FAR. FAR describes how many of the predicted ``yes" events were false alarms, however, it ignores misses and consequently has to be used in conjunction with POD. It ranges from 0 to 1, with perfect score FAR\,=\,0 (no false alarms). BIAS measures the ratio of the frequency of forecasts to the frequency of observations and ranges from 0 to $\infty$, with perfect score BIAS\,=\,1. It reveals whether the forecast has a tendency to underforecast (BIAS$<$1) or overforecast (BIAS$>$1) events. However, it tells nothing about how well the forecast corresponds to the observations. Finally, HSS estimates the accuracy of the forecast relative to that of random chance. It ranges from -$\infty$ to 1, where HSS\,=\,1 is a perfect score, HSS\,=\,0 means that the forecast is no better than random, and HSS$<$0 means that the forecast is worse than random. The number of possible outcomes based on the contingency table, as well as the corresponding verification measures for both the training and the test set, are given in Table \ref{tab5} for the following ``events": $k=0$ ($|FD|<1\,\%$), $k=1$ ($1\,\%<|FD|<3\,\%$), $k=2$ ($3\,\%<|FD|<6\,\%$), $k=3$ ($|FD|>6\,\%$), $k=0,1,2$ ($|FD|<6\,\%$), $k=0,1$ ($|FD|<3\,\%$), $k=1,2,3$ ($|FD|>1\,\%$). We note that the first four ``events" correspond to the four bins of the probability distribution in Sections \ref{statistics} and \ref{model}, whereas the last three ``events" represent a less specific forecast. For these two groups of events verification measures are also presented separately in Figure \ref{fig7} for the training and test sets.

%

%
\begin{table}
\caption{The number of possible outcomes based on the contingency table and the corresponding verification measures for validation and test sets, for different events}
\label{tab5}
\begin{tabular}{cccccccccc} 
						& event		& a		& b		& c		& d		& POD		& FAR		& BIAS	& HSS\\
\hline
	          & $k=0$	  & 45	& 22	& 34	& 86	& 0.57	& 0.33	& 0.85	& 0.37\\
	          & $k=1$	  & 21	& 34	& 42	& 90	& 0.33	& 0.62	& 0.87	& 0.06\\
training  	& $k=2$	  & 8	  & 25	& 21	& 133	& 0.28	& 0.76	& 1.14	& 0.11\\
set       	& $k=3$	  & 10	& 22	& 6	  & 149	& 0.63	& 0.69	& 2.00	& 0.34\\
\cline{2-10}
   	        & $k<3$	  & 149	& 6	  & 22	& 10	& 0.87	& 0.04	& 0.91	& 0.34\\
	          & $k<2$	  & 110	& 12	& 32	& 33	& 0.77	& 0.10	& 0.86	& 0.44\\
	          & $k>0$	  & 86	& 34	& 22	& 45	& 0.80	& 0.28	& 1.11	& 0.37\\
\hline
	          & $k=0$	  & 8	  & 6	  & 9	  & 19	& 0.47	& 0.43	& 0.82	& 0.24\\
	          & $k=1$	  & 3	  & 11	& 4	  & 24	& 0.43	& 0.79	& 2.00	& 0.08\\
test	      & $k=2$	  & 3	  & 4	  & 8	  & 27	& 0.27	& 0.57	& 0.64	& 0.16\\
set	        & $k=3$	  & 3	  & 4	  & 4	  & 31	& 0.43	& 0.57	& 1.00	& 0.31\\
\cline{2-10}
	          & $k<3$	  & 31	& 4	  & 4	  & 3	  & 0.89	& 0.11	& 1.00	& 0.31\\
	          & $k<2$	  & 21	& 7	  & 3	  & 11	& 0.88	& 0.25	& 1.17	& 0.50\\
	          & $k>0$	  & 19	& 9	  & 6	  & 8	  & 0.76	& 0.32	& 1.12	& 0.24\\
\end{tabular}
\end{table}

It can be seen from Table \ref{tab5} and Figure \ref{fig7} that there are differences in the verification measures between the training and test sets, especially for the BIAS in case of a more specific forecast (\textit{i.e.} for the forecast of a specific bin $[k=0,1,2,3]$). However, the differences are not pronounced and moreover they are not systematic, indicating that the successfulness of the forecast mainly relies on the approximations used and not the sample. The forecast of the intermediate bins $k=1,2$ is least reliable, since we get the lowest number of hits and largest number of false alarms. This is also evident in the HSS, which gives lowest values indicating that the forecast is only slightly better than random for these two bins. This indicates that the forecast has a ``resolution" problem, \textit{i.e.} it has difficulties in discerning between neighboring bins. This is also supported by the fact that when less specific bins are regarded ($k<3$, $k<2$, and $k>0$), POD is much higher, FAR is lower, BIAS is closer to perfect value (BIAS$\,\approx\,1$) and HSS has larger positive values, the latter indicating that the forecast shows skill compared to random forecast (see Table \ref{tab5} and Figures \ref{fig7}c and \ref{fig7}d). Therefore, we conclude that the Forbush decrease prediction is more reliable for a less specific forecast, \textit{i.e.} for predicting whether or not CME will be GCR-effective ($k>0 \leftrightarrow FD>1\,\%$), whether or not it will be strongly/intensly GCR-effective ($k<2 \leftrightarrow FD<3\,\%$) and whether or not it will be intensly GCR-effective ($k<3 \leftrightarrow FD<6\,\%$). Given the verification measures presented in Table \ref{tab5} and Figure \ref{fig7}, the most reliable forecast (highest POD, lowest FAR, BIAS\,$\approx\,1$, and high HSS) is the prediction whether or not CME will be strongly/intensly GCR-effective ($k<2 \leftrightarrow FD<3\,\%$), \textit{i.e.} predicting whether CME will produce $FD>3\,\%$.

\section{Summary and conclusions}
\label{conclusion}

We used a sample of CME--flare pairs detected remotely, and the associated cosmic ray (CR) count levels at Earth for the purpose of forecasting the CME-associated Forbush decrease (FD) magnitude $[FD]$. The advantage of the approach is in the early forecast, since the travel time for a CME from Sun to Earth is of the order of $\approx$ one day. To characterize CME/flare event we use L1 coronagraphic CME observations, the EUV flare-position observation, as well as the soft X-ray flare measurements. We note that some of the properties derived from these observations can also be obtained from ground-based measurements (\textit{e.g.} proxy of the CME speed can be obtained from solar Type-II radio bursts, flare position can be obtained by H\, $\alpha$ observations). Therefore, the remotely observed CME/flare properties are not necessarily spacecraft-dependent.

The relationship between FD magnitude at the Earth and remote observations of CMEs and associated solar flares is studied \textit{via} statistical analysis. It was found that FD magnitude is larger for faster CMEs with larger apparent width, associated with stronger flares, originating close to the center of the solar disc and (possibly) involved in a CME--CME interaction. These relations are quantified through the change in the distribution of FD magnitude, which is mathematically reconstructed using the shifted geometric distribution. The reconstructed distributions are used to obtain a joint probability distribution for a certain CME/flare event, where we use the sample of 187 CME--flare--FD associations as an ensemble of possibilities for a certain event. The joint probability distribution for a certain CME/flare event behaves differently when different CME/flare properties are used as input, reflecting the behavior found by statistical analysis. However, distributions are always highly asymmetric with the greatest probability that CME will not be GCR-effective, which is the general behavior of CMEs (a large majority of CMEs will never reach the Earth and/or will not be very GCR-effective). Therefore, we impose empirically optimized thresholds on the probability distribution to obtain the estimation of the GCR-effectiveness for a specific CME/flare event. In this way we obtain an empirical probabilistic model that uses selected remote solar observations of CME and associated solar flare as an input and gives expected FD magnitude range as an output.

Evaluation of the forecast method is performed on the training set (the sample of 187 CME--flare--FD associations used for the statistical analysis) and test set (independent sample of 42 CME--flare--FD associations). The evaluation revealed that the forecast is less reliable when it is more specific, due to difficulties in discerning between neighboring bins. It was found especially ineffective for prediction of intermediate FD magnitudes ($1\,\%<FD<3\,\%$ and $3\,\%<FD<6\,\%$). However, when the forecast is less specific, the quality of the forecast improves. Based on the performed evaluation, the Forbush-decrease prediction is found to be most reliable in predicting whether or not CME will produce $FD>3\,\%$.

Based on the research presented in this study, an online application for the prediction of Forbush decrease magnitude based on the remote solar observations of CMEs and associated solar flares, ``Forbush Decrease Forecast Tool (FDFT)" was developed and is publically availabale at \url{oh.geof.unizg.hr/FDFT/fdft.php}. The full training-set list, as well as the test set list are also available at the same webpage under ``Documentation".

%

%
\begin{acks}
This work has been supported in part by Croatian Science Foundation under the project 6212 ``Solar and Stellar Variability". M. Dumbovi\' c and J. \v Calogovi\' c acknowledge the support by the ESF project PoKRet.
\end{acks}

%
\section*{Disclosure of Potential Conflicts of Interest}
The authors declare that they have no conflicts of interest.

%
%
\bibliographystyle{spr-mp-sola}
\bibliography{REFs}

\end{article} 
\end{document}